\documentclass[twocolumn]{aastex63}

\begin{document}

\title{Constraining hadron-quark phase transition parameters within the quark-mean-field model using multimessenger observations of neutron stars}

%\correspondingauthor{Ang Li}
%\email{liang@xmu.edu.cn}

\author{Zhiqiang Miao}
\affiliation{Department of Astronomy, Xiamen University, Xiamen, Fujian 361005, China}

\author{Ang Li}
\affiliation{Department of Astronomy, Xiamen University, Xiamen, Fujian 361005, China}

\author{Zhenyu Zhu}
\affiliation{Department of Astronomy, Xiamen University, Xiamen, Fujian 361005, China}
\affiliation{Institute for Theoretical Physics, Frankfurt am Main 60438, Germany}

\author{Sophia Han}
\affiliation{Department of Physics and Astronomy, Ohio University, Athens, OH 45701, USA}
\affiliation{Department of Physics, University of California Berkeley, Berkeley, CA 94720, USA}

\date{\today}

\begin{abstract}
We extend the quark mean-field (QMF) model for nuclear matter and study the possible presence of quark matter inside the cores of neutron stars. 
A sharp first-order hadron-quark phase transition is implemented combining the QMF for the hadronic phase with ``constant-speed-of-sound'' parametrization for the high-density quark phase. 
The interplay of the nuclear symmetry energy slope parameter, $L$, and the dimensionless phase transition parameters (the transition density $n_{\rm trans}/n_0$, the transition strength $\Delta\varepsilon/\varepsilon_{\rm trans}$, and the sound speed squared in quark matter $c^2_{\rm QM}$) are then systematically explored for the hybrid star proprieties, especially the maximum mass $M_{\rm max}$ and the radius and the tidal deformability of a typical $1.4 \,M_{\odot}$ star.
We show the strong correlation between the symmetry energy slope $L$ and the typical stellar radius $R_{1.4}$, similar to that previously found for neutron stars without a phase transition. 
With the inclusion of phase transition, we obtain robust limits on the maximum mass ($M_{\rm max}< 3.6 \,M_{\odot}$) and the radius of $1.4 \,M_{\odot}$ stars ($R_{1.4}\gtrsim 9.6~\rm km$), and 
we find that a too-weak ($\Delta\varepsilon/\varepsilon_{\rm trans}\lesssim 0.2$) phase transition taking place at low densities $\lesssim 1.3-1.5 \, n_0$ is strongly disfavored.
We also demonstrate that future measurements of the radius and tidal deformability of $\sim 1.4 \,M_{\odot}$ stars, as well as the mass measurement of very massive pulsars, can help reveal the presence and amount of quark matter in compact objects. 
\end{abstract}


%Unified Astronomy Thesaurus concepts: Neutron star cores (1107); Neutron stars (1108); Gravitational waves (678)

%--------|---------|---------|---------|---------|---------|---------|---------|
\section{Introduction}

The nature of matter under extreme conditions of density, pressure (gravity), isospin and magnetic field accessible only in the dense cores of neutron stars still remains an open question. In particular, the mass and radius of neutron stars encode unique information on the equation of state (EOS) at supranuclear densities.
Several massive pulsars with a mass about two-solar masses detected during the last decade set stringent constraints on EOS of neutron star matter, PSR J1614-2230 ($M=1.908\pm 0.016 \,M_{\odot}$) ~\citep{2010Natur.467.1081D,2016ApJ...832..167F,2018ApJS..235...37A}, PSR J0348+0432 ($M=2.01\pm 0.04\, M_{\odot}$)~\citep{2013Sci...340..448A}, and MSP J0740+6620 ($M=2.14^{+0.10}_{-0.09} \,M_{\odot}$)~\citep{2020NatAs...4...72C}, for which masses are reported with %$1\sigma$ error-bars or equivalently 
$68.3\%$ credibility intervals, respectively.
There has been a simultaneous estimation of the mass and radius of neutron stars by the NASA Neutron Star Interior Composition ExploreR (NICER) mission from pulse-profile modeling of accretion hot spots of the isolated millisecond pulsar PSR J0030+0451~\citep{2019ApJ...887L..24M,2019ApJ...887L..22R,2019ApJ...887L..21R}, $M=1.44^{+0.15}_{-0.14} \,M_{\odot}$, $R=13.02^{+1.24}_{-1.06}~\rm km$~\citep{2019ApJ...887L..24M}
and $M=1.34^{+0.15}_{-0.16} \,M_{\odot}$, $R=12.71^{+1.14}_{-1.19}~\rm km$~\citep{2019ApJ...887L..21R}, to the $68.3\%$ credibility interval. 
The detection of the GW170817 binary neutron star merger event~\citep{2017PhRvL.119p1101A} with its electromagnetic counterpart has also greatly advanced the study of dense matter at extreme densities~\citep[e.g.,][]{2017ApJ...850L..34B,2017ApJ...850L..19M,2018PhRvL.121p1101A,2018PhRvL.120q2703A,2018ApJ...852L..29R,2018ApJ...852L..25R,2018PhRvD..97b1501R,2018PhRvD..97h3015Z,2018ApJ...862...98Z,2019PhRvX...9a1001A,2019PrPNP.10903714B,2019PhRvL.123n1101F,2019JPhG...46l3002G,2019ApJ...878..159M,2019PhRvD.100b3015S,2019ApJ...881...73W,2019PhRvD..99l1301Z,2020ApJ...893..146A,2020arXiv200400846B,2020NatAs...4..625C,2020PhRvD.101f3007E,2020ApJ...892...55J,2020PhRvD.101j3021O,2020ApJ...893L..21R,2020ApJS..250....6W,2020arXiv200502677Z}.
By constructing the neutron star EOS using chiral effective field theory of neutron matter and combining with multimessenger observations of GW170817, \citet{2020NatAs...4..625C} found that the radius of a $1.4 \,M_{\odot}$ neutron star is $R_{1.4}= 11.0^{+0.9}_{-0.6}~\rm km$ ($90 \%$ credible interval) assuming that a description in terms of nucleonic degrees of freedom remains valid up to $2\,n_0$, where $ n_0 = 0.16~\rm fm^{-3}$ is the nuclear saturation density.

In recent years much attention has been paid to one of the main features of the EOS, i.e. the symmetry energy~\citep[e.g.,][]{2014EPJA...50....9L,2016PrPNP..91..203B,2017RvMP...89a5007O}.
The behavior of the nuclear symmetry energy as a function of density is crucial for interpreting many astrophysical observations related to compact objects, including the overall structure of neutron stars~\citep[e.g.,][]{2018ApJ...862...98Z,2019JPhG...46g4001K,2019PhRvC.100c5801P,2019ApJ...883..174X,2019ApJ...885..121R,2019PhRvC..99b5804Z,2020arXiv200407232D}.
It has been shown that it is possible to use the observation of the global properties of neutron stars to put constraints on the symmetry energy (especially its slope with respect to the density) at saturation~\citep[e.g.,][]{2001ApJ...550..426L,2006PhLB..642..436L,2013ApJ...773...11H, 2014EPJA...50...40L}.

At high densities reached in the interiors of massive neutron stars (possibly up to $\approx 8-10\,n_0$), quark degrees of freedom may start appearing and play a role. 
The possible appearance of quark matter and deconfinement phase transition is one of 
unresolved puzzles of neutron star matter and could largely affect the underlying EOS. 
At present, since the quark matter EOS is poorly known at zero temperature and high density appropriate for neutron stars, one possible way of tackling the problem is to perform the calculations with certain quark matter models in sufficient large parameter space and then compare the predictions with observations of neutron star static and dynamical properties, which has been of special interest in the present era of gravitational wave astronomy~\citep[e.g.,][]{2018ApJ...860..139B,2018ApJ...857...12N,2018PhRvD..97h4038P,2019MNRAS.484.4980A,2019PhRvL.122f1102B,2019PhRvD..99b3009C,2019ApJ...877..139G,2019PhRvD..99h3014H,2019PhRvD.100j3022H,2019PhRvD..99j3009M,2019PhRvL.122f1101M,2019JPhG...46g3002O,2019A&A...622A.174S,2020PhRvD.101d4019C,2020ApJ...893L...4C,2020PhRvD.101f3007E, 2020arXiv200409566M,2020ApJ...896..109N,2020ApJ...895...28P,2020PhRvD.101l3029T,2020PhRvL.124q1103W}.

When extracting dense matter properties from observations, it is difficult to eliminate the model dependence considering a large sample of nuclear matter EOS models, since there can be more than one physical quantity from the theoretical input which the neutron star observables are sensitive to. 
Alternatively, one can construct theoretical EOSs that satisfy the same criterion for other quantities, for instance with other saturation properties fixed, and decouple the dependence on nuclear symmetry energy slope explicitly~\citep[see e.g.,][]{2018ApJ...862...98Z,2019PhRvC..99b5804Z,2020arXiv200705116L}.
In this paper, we discuss the previously proposed quark mean-field (QMF) model~\citep{1998PhRvC..58.3749T} which allows one to tune the density dependence of the symmetry energy in a self-consistent way~\citep{2018PhRvC..97c5805Z,2018ApJ...862...98Z,2019PhRvC..99b5804Z}, in combination with the constant-speed-of-sound (CSS) parametrization for high-density quark matter EOS~\citep{2013PhRvD..88h3013A}. 
We perform calculations of the mass-radius relation and tidal deformability for normal hadronic  and hybrid star configurations, using various choices of the symmetry energy slope parameter and the hadron-quark phase transition parameters. 
Moreover, we examine possible correlations among the symmetry energy slope, the neutron star maximum mass, and the radius of a canonical $1.4 \,M_{\odot}$ star, and the tidal deformabilities deduced from GW170817-like events.

This paper is organized as follows. In Sec.~\ref{sec:qmf} we discuss the EOS for the hadronic phase of neutron stars, i.e., the QMF model; in Sec.~\ref{sec:css} we apply the CSS parametrization to describe the quark phase. In Sec.~\ref{sec:l} we mainly discuss effects from the symmetry energy slope on the mass, radius, and tidal deformability of hybrid stars, and then confront the results of our calculations with multimessenger observations in Sec.~\ref{sec:14}.
Finally we summarize in Sec.~\ref{sec:sum}.

%--------|---------|---------|---------|---------|---------|---------|---------|
\section{Nuclear matter within the QMF} 
\label{sec:qmf}

%____________________________________
\begin{table*}[htb]
\centering
\caption{Saturation properties used for the fitting of meson coupling parameters ($g_{\sigma q}, g_{\omega q}, g_{\rho q}, g_2, g_3, \Lambda_v$) in the QMF Lagrangian [Eq.~(\ref{eq:lag})]: The saturation density $n_0$ (in fm$^{-3}$) and corresponding values at the saturation point for the binding energy $E/A$ (in MeV), the incompressibility $K$ (in MeV), the symmetry energy $E_{\rm sym}$ (in MeV), the symmetry energy slope $L$ (in MeV) and the ratio between the effective mass and free nucleon mass $M_N^\ast/M_N$. The corresponding empirical data~\citep{2006EPJA...30...23S,2012ChPhC..36....3M,2013ApJ...771...51L,2017RvMP...89a5007O} are also collected.}
\vskip-4mm\
\setlength{\tabcolsep}{17.5pt}
\renewcommand{\arraystretch}{1.1}
\begin{tabular}{c|cccccc}\hline\hline
& $\rho_0$ & $E/A$  & $K$ & $E_{\rm sym}$& $L$ & $M_N^\ast/M_N$ \\ 
& $[{\rm fm}^{-3}]$ & [MeV] & [MeV] & [MeV] & [MeV] & / \\ \hline
QMF & 0.16 & -16.0 & 240 & 31 &$30-60$ & 0.77 \\ 
Exp. & $0.16\pm0.01$ & $-16.0\pm1.0$ & $240\pm20$ & $31.7\pm3.2$ & $\approx30-86$ & $\approx0.6-1$ \\\hline\hline
\end{tabular} \label{tab:sat}
\vskip -4mm\
\end{table*}

The EOS of nuclear matter obtained within the QMF model has been amply discussed in previous works~\citep[e.g.,][]{2018PhRvC..97c5805Z,2018ApJ...862...98Z,2019PhRvC..99b5804Z}. For a review see \citet{2020arXiv200705116L}. We first adopt a harmonic oscillator potential to confine quarks in a nucleon, with its parameters determined by the mass and radius of free nucleon, and then connect the nucleon in the medium with a system of many nucleons which interact through exchanging $\sigma,\omega$, and $\rho$ mesons. 
The Lagrangian in the mean-field approximation can be written as:
\begin{widetext}
\vskip-8mm\
\begin{eqnarray}\label{eq:lag}
 \mathcal{L} &=&  \overline{\psi}\left(i\gamma_\mu \partial^\mu - M_N^\ast  
 - g_{\omega N}\omega\gamma^0 - g_{\rho N}\rho\tau_{3}\gamma^0\right)\psi \nonumber \\
& & 
- \frac{1}{2}(\nabla\sigma)^2 - \frac{1}{2}m_\sigma^2 \sigma^2 - \frac{1}{3} g_2\sigma^3 - \frac{1}{4}g_3\sigma^4 \nonumber + \frac{1}{2}(\nabla\rho)^2+ \frac{1}{2}(\nabla\omega)^2 %\\
%& & 
+ \frac{1}{2}m_\omega^2\omega^2 
 + \frac{1}{2}m_\rho^2\rho^2 + \frac{1}{2}g_{\rho N}^2\rho^2 \Lambda_v g_{\omega N}^2\omega^2,
\end{eqnarray}
\end{widetext}
where $g_{\omega N}$ and $g_{\rho N}$ are the nucleon coupling constants for $\omega$ and $\rho$ mesons, 
$g_{\sigma q}$, $g_{\omega q}$, and $g_{\rho q}$ are the coupling constants of $\sigma$, $\omega$, and $\rho$ mesons with quarks, respectively.
$m_{\sigma} = 510~\rm{MeV}$,~$m_{\omega}=783~\rm{MeV}$, and $m_{\rho}=770~\rm{MeV}$ are the meson masses. 
From the quark counting rule, we obtain $g_{\omega N}=3g_{\omega q}$ and $g_{\rho N}=g_{\rho q}$. 
The calculation of the confined quarks in a nucleon gives rise to the effective quark mass $m_q^\ast = m_q-g_{\sigma q}\sigma$, as well as the relation of the effective nucleon mass $M_N^*$ as a function of the $\sigma$ field, $g_{\sigma N} = -\partial M_N^\ast/\partial \sigma$~\citep[e.g.,][]{1998PhRvC..58.3749T,2000PhRvC..61d5205S}.
It is noteworthy that, compared to the constant coupling in the standard Walecka model, the coupling treatment in QMF is consistently generated from the confined quark description. This difference in $\sigma$-nucleon coupling results in the main distinction of QMF from other mean-field models.
The cross coupling from $\omega$ meson and $\rho$ meson, $\frac{1}{2}g_{\rho N}^2\rho^2 \Lambda_v g_{\omega N}^2\omega^2$, can largely improve the descriptions on the symmetry energy $E_{\rm sym}(n)$ and give a reasonable value of the symmetry energy slope $L$~\citep[e.g.,][]{2001PhRvL..86.5647H,2018PhRvC..97c5805Z}. 

The energy density $\varepsilon$ of nuclear matter is generated from the energy-momentum tensor related to the QMF Lagrangian [Eq.~(\ref{eq:lag})], as a function of the relevant partial densities $n_i$ ($i=n,p$).
The parabolic approximation is usually applicable, and the energy per nucleon can be written as [$\beta\equiv(n_n-n_p)/n$] 
\begin{eqnarray}
E/A (n, \beta) \approx  E/A (n, \beta=0) + E_{\rm sym}(n)\beta^2, \label{eq:ba}
\end{eqnarray}
and $E/A (n, \beta=0)$ can be expanded around the saturation density $n_0$
\begin{eqnarray}
E/A (n,0) = E/A (n_0) +\frac{1}{18} K\frac{n-n_0}{n_0}, \label{eq:ba0}
\end{eqnarray}
with $K$ the incompressibility at the saturation point.
The symmetry energy $E_{\rm sym}(n)$ is expressed in terms of the difference of the energy per particle between pure neutron ($\beta=1$) matter and symmetric ($\beta=0$) nuclear matter, $E_{\rm sym}(n)\approx E/A (n, 1) -E/A (n, 0)$. In order to characterize its density dependence, $E_{\rm sym}(n)$ can be expanded around the saturation density $n_0$ as follows
\begin{eqnarray}
E_{\rm sym}(n) &= E_{\rm sym}(n_0)  + \frac{dE_{\rm sym}}{dn}(n-n_0)  + ...   \label{eq:esym0}
\end{eqnarray}
and the following parameters can be defined (both having an energy dimension (MeV))
\begin{eqnarray}
E_{\rm sym} = E_{\rm sym}(n_0),~L = 3n_0 \left(\frac{dE_{\rm sym}}{dn}\right)_{n_0}.
\end{eqnarray}
Other thermodynamical quantities can also be obtained including the chemical potential and pressure
\begin{eqnarray}\label{eq:mu}
\mu_i&=&\frac{\partial \varepsilon}{\partial n_i},\\
p(n)&=& n^2\frac{d}{dn}\frac{\varepsilon}{n}=n\frac{d\varepsilon}{dn}-\varepsilon=n\mu_B-\varepsilon.
\end{eqnarray}

In the present study, we employ the parameter sets ($g_{\sigma q}, g_{\omega q}, g_{\rho q}, g_2, g_3, \Lambda_v$) previously fitted in \citet{2018ApJ...862...98Z} from reproducing the empirical saturation properties of nuclear matter: the saturation density $n_0$ and corresponding values at saturation point for the binding energy $E/A$, the incompressibility $K$, the symmetry energy $E_{\rm sym}$, the symmetry energy slope $L$, and the effective mass $M_N^\ast$. The values employed together with corresponding empirical ones are collected in Table \ref{tab:sat}. While the EOS of symmetric nuclear matter has been relatively well-constrained~\citep{2002Sci...298.1592D}, matter with nonzero isospin asymmetry remains unknown, largely due to the uncertainty in the symmetry energy~\citep[e.g.,][]{2014EPJA...50....9L}.
The symmetry energy slope $L$ characterizes the density dependence of $E_{\rm sym}$, and is one of the key nuclear parameters that dominate the ambiguity and stiffness of EOS for dense matter at higher densities in the absence of phase transition with strangeness. Therefore, as shown in Table \ref{tab:sat}, we choose values of $L$ in its empirical range~\citep[e.g.,][]{2013PhLB..727..276L,2013ApJ...771...51L,2014NuPhA.922....1D,2017RvMP...89a5007O} as input of the parameter fitting, and study its effect on the properties of (binary) neutron stars.
The upper bound of $\approx60$ MeV is also consistent with prediction from the unitary gas conjecture ~\citep{2017ApJ...848..105T}, with $E_{\rm sym} $ at saturation being set to its preferred value of $31$ MeV~\citep[e.g.,][]{2013PhLB..727..276L,2014NuPhA.922....1D,2017RvMP...89a5007O}. 
This independent constraint on $L$ is to ensure that neutron matter energy is larger than the unitary gas energy at low densities $\lesssim 1.5\,n_0$. We mention here that the present study has neglected higher-order expansion terms in the energy densities (Eqs.~\ref{eq:ba}-\ref{eq:esym0}), which may become important for dense neutron-rich matter~[see e.g., \citet{2018PhRvC..98c5804M,2020arXiv200803469M,2020arXiv200811338L,2019ApJ...879...99Z,2019EPJA...55...39Z,2020arXiv200203210Z} for some latest discussions on higher order terms], however, it is not guaranteed that at high enough densities nucleonic degrees of freedom will still dominate.

%--------|---------|---------|---------|---------|---------|---------|---------|
\section{High-density matter with the CSS parametrization} 
\label{sec:css}

For the high-density quark phase we utilize the CSS parametrization~\citep{2013PhRvD..88h3013A}, making use of the feature that for a considerable class of microscopic quark matter models the speed of sound turns out weakly density-dependent, for example the Nambu-Jona--Lasinio (NJL) model~\citep{2010PhRvD..81b3009A,2013A&A...551A..61Z,2016PhRvC..93d5812R}, field correlator method~\citep{2015PhRvD..92h3002A}, variations of the MIT bag model~\citep{1976PhLB...62..241B}, et cetera. Assuming the sound speed in quark matter is density independent from the first-order transition onset up to the maximum central pressure of a star, the CSS parametrization is applicable to high-density EOSs for which there is a sharp interface (Maxwell construction) between bulk hadronic matter and quark matter, i.e. the quark-hadron surface tension is high enough to disfavor mixed phases (Gibbs construction). It has been shown that strong first-order phase transition with a sharp interface is the most promising scenario to be tested or distinguished from pure hadronic matter by future observations~\citep{2019PhRvD..99h3014H,2019PhRvD.100j3022H,2020PhRvD.101d4019C}. One can also formulate EOSs that model quark-hadron interfaces which are mixed~\citep[e.g.,][]{2008IJMPE..17.1635L,2011PhRvC..83b5804B,2015PhRvC..91c5803L,2020PhRvD.101l3030F,2020arXiv200409566M} or feature a smooth crossover~\citep[e.g.,][]{2019ApJ...885...42B}, given current uncertainties regarding the nature of the phase transition.

%fig
\begin{figure}%[htb]
\centering
\includegraphics[width=20pc]{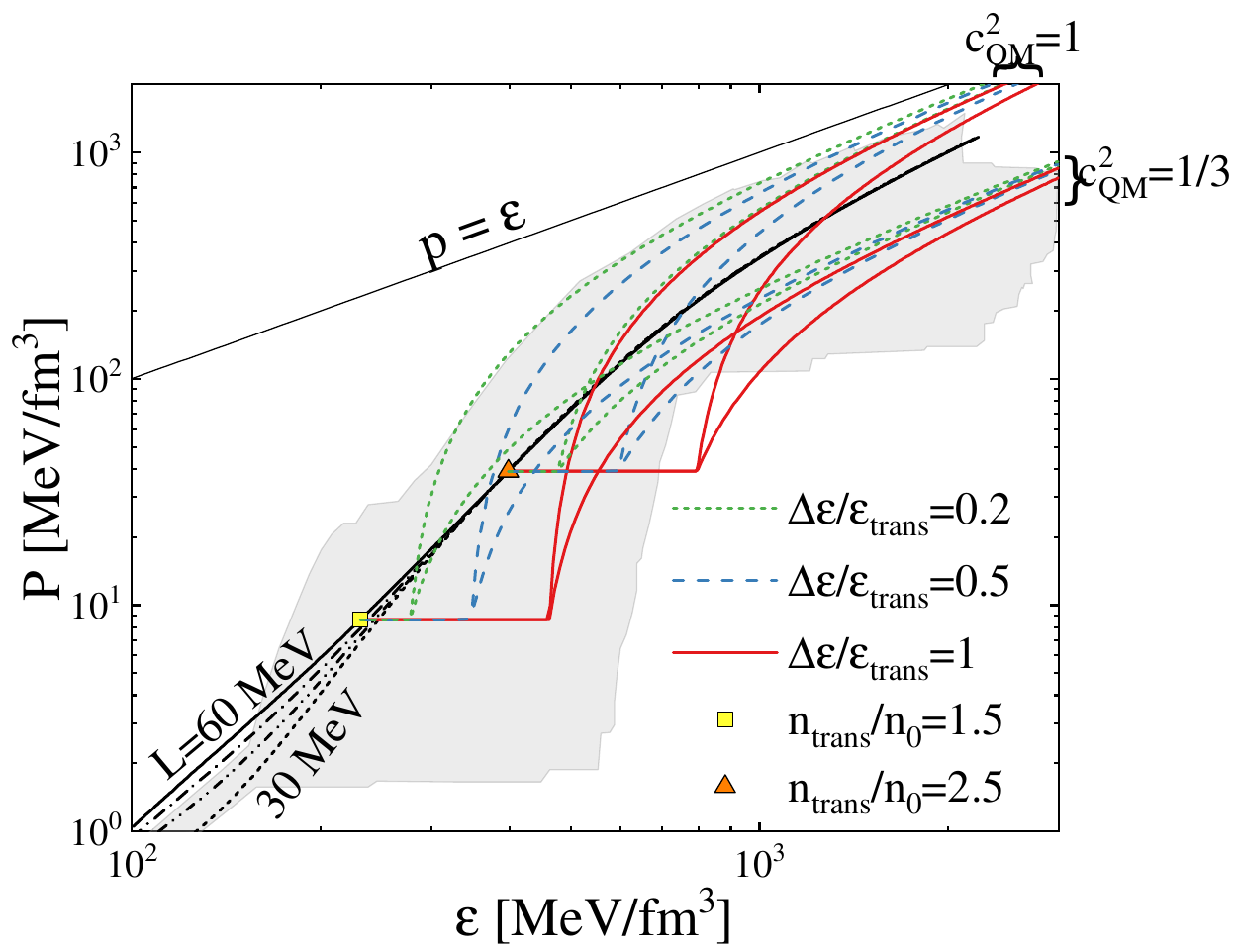}
\caption{Exemplary hybrid EOSs (colored curves) with a sharp first-order phase transition from hadronic matter (QMF, $L=60~\rm MeV$) to quark matter (CSS), at different transition densities $n_{\rm trans}/n_0=1.5,~2.5$ with different transition strengths $\Delta\varepsilon/\varepsilon_{\rm trans}=0.2,~0.5,~1$. Two groups of colored curves represent the causality limit $c^2_{\rm QM} =1$ and the conformal limit $c^2_{\rm QM} =1/3$, respectively, in the high-density phase. The QMF results for normal neutron stars with $L=30,~40,~50~\rm MeV$ are also shown for comparison (black curves).  
The shaded background is the generic family from the maximal model \citep{2019EPJA...55...97T} constrained at low densities by state-of-the-art calculations of neutron-rich matter from chiral effective field theory, allowing the complete parameter space for the speed of sound above $n=n_0$ that is compatible with the LIGO/Virgo constraint from GW170817 ($70\leq\tilde{\Lambda}\leq 720$)~\citep{2019PhRvX...9a1001A}. Extreme causal EOS is also shown with the straight solid line.
} 
\label{fig:eos}
\end{figure}

The dimensionless CSS parameters are the squared speed of sound in the high-density phase $c_{\rm QM}^2$ (we work in units where $\hbar=c=1$), the hadron-quark phase transition density $n_{\rm trans}/n_0$, and the discontinuity in the energy density at the transition $\Delta\varepsilon/\varepsilon_{\rm trans}$ where $n_{\rm trans} \equiv n_{\rm HM}(p_{\rm trans})$ and $\varepsilon_{\rm trans} \equiv \varepsilon_{\rm HM}(p_{\rm trans})$. For a given hadronic matter EOS $\varepsilon_{\rm HM}(p)$, the full EOS is 
\begin{eqnarray}
\varepsilon(p) = \left\{\!
\begin{array}{ll}
\varepsilon_{\rm HM}(p), & p<p_{\rm trans} \\ \nonumber 
\varepsilon_{\rm HM}(p_{\rm trans})+\Delta\varepsilon+c_{\rm QM}^{-2} (p-p_{\rm trans}), & p>p_{\rm trans}  
\end{array}
\right.
\end{eqnarray}
The high-pressure CSS EOS can be written as \citep{2013PhRvD..88h3013A,2013A&A...551A..61Z},
\begin{eqnarray}
p(\mu_B) &=& A\mu_B^{1+1/c_{\rm QM}^2}-B\label{eq:css}\\
\mu_B (p) &=& [(p+B)/A]^{c_{\rm QM}^2/(1+c_{\rm QM}^2)}\\
n(\mu_B) &=& (1+1/c_{\rm QM}^2)A\mu_B^{1/c_{\rm QM}^2} 
\end{eqnarray}
where $A$ is a parameter with energy dimension $3-c_{\rm QM}^{-2}$ and $B=(\varepsilon_{\rm trans}+\Delta\varepsilon-c_{\rm QM}^{-2}\,p_{\rm trans})/(1+c_{\rm QM}^{-2})$.
To construct a first-order transition from some low-pressure
EOS to a high-pressure EOS of Eq.~(\ref{eq:css}), $A$ is chosen such that the pressure is monotonically increasing with  $\mu_B$ and the baryon number density does not decrease at the transition. 

%fig
\begin{figure*}[htb]
\centering
\includegraphics[width=21.5pc]{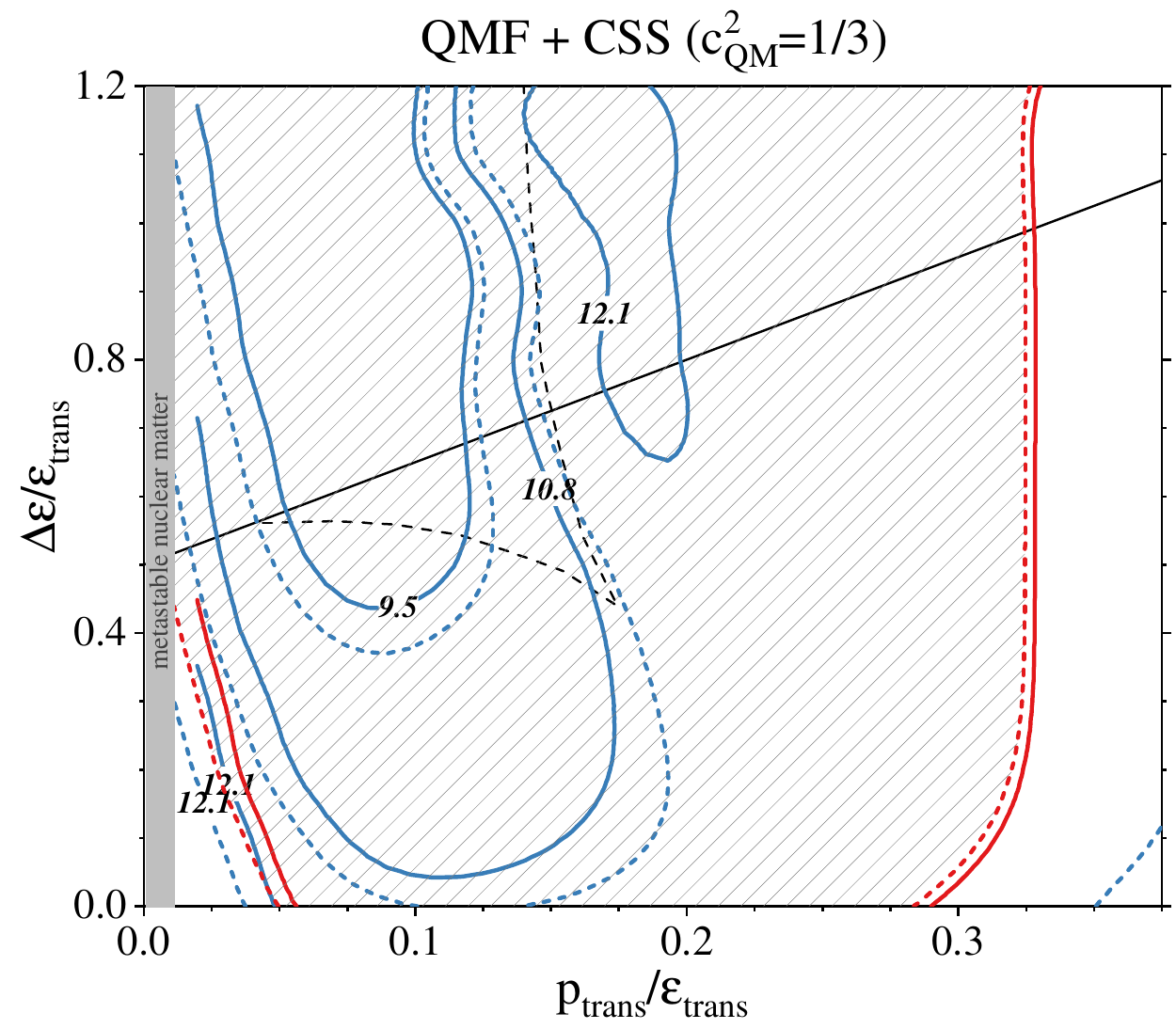}
\includegraphics[width=20.5pc]{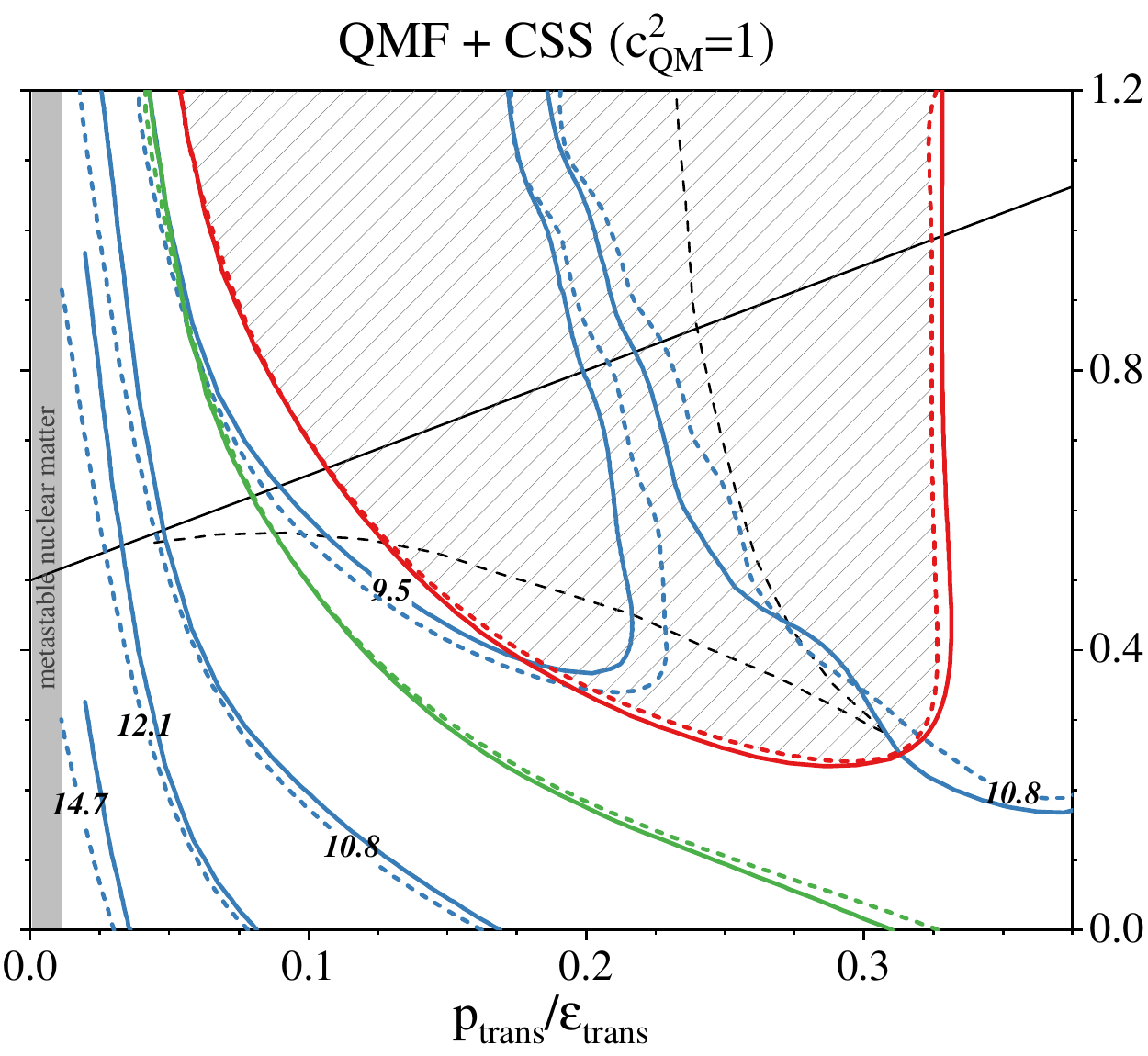}
\caption{\label{fig:rcontour} Contour plots for the maximum mass of hybrid stars $M_{\rm max}$ (red) and the radius of maximum-mass stars $R_{\rm max}$ (blue) as a function of the CSS parameters in high-density phase. Each panel shows the dependence on the CSS parameters ($n_{\rm trans}/n_0,\Delta\varepsilon/\varepsilon_{\rm trans}$) at fixed quark matter sound speed $c^2_{\rm QM} =1/3$ (left) and $c^2_{\rm QM} =1$ (right). Both results with $L=30$ MeV (dashed curves) and $L=60$ MeV (solid curves) are shown. Low pressure/density regions where $n_{\rm trans} < n_0$ are excluded, with the (leftmost) grey-shaded band showing the $L=30$ MeV case (for $L=60$ MeV the band is slightly extended to the right); the highest pressure/density reached in the center of the heaviest hadronic star within QMF is $p_{\rm cent}/\varepsilon_{\rm cent}\simeq0.43$.
The hatched regions inside the $2\,M_{\odot}$ contours are excluded by $\gtrsim2\,M_{\odot}$ pulsars observed~\citep{2010Natur.467.1081D,2013Sci...340..448A,2016ApJ...832..167F,2018ApJS..235...37A,2020NatAs...4...72C}. $2.14\,M_{\odot}$ contours (green) are also shown in the right panel, reflecting the intermediate value of the heaviest pulsar recently discovered MSP J0740+6620~\citep{2020NatAs...4...72C}. The solid black line denotes the threshold value $\Delta\varepsilon_{\rm crit}$ [Eq.(\ref{eq:limit})] below which there is always a stable hybrid star branch connected to the hadronic branch. The dashed black lines mark the border of regions where the disconnected hybrid star branch exists.}
\label{fig:rcontour}
\end{figure*}

We perform calculations by varying $c_{\rm QM}^2$ from the causality limit $c_{\rm QM}^2=1$ to the conformal limit $c_{\rm QM}^2=1/3$ (the value for systems with conformal symmetry that may be applicable to relativistic quarks). It is worth mentioning that perturbative QCD calculations exhibit quark matter with $c_{\rm QM}^2$ around 0.2 to 0.3~\citep{2010PhRvD..81j5021K}; see detailed analysis of the sound speed behaviour in dense matter~\citep[e.g.,][]{2015PhRvL.114c1103B,2019arXiv190600826X}.

We illustrate in Fig.~\ref{fig:eos} the EOSs $P(\varepsilon)$ for dense matter with sharp first-order phase transitions applying the generic CSS parametrization for quark matter, specified by the three CSS parameters $n_{\rm trans}$, $\Delta\varepsilon$, and $c_{\rm QM}^2$.
For comparison, we also show the causality limit $P=\varepsilon$ (thin straight line) together with baseline EOSs (shaded background) from the maximal model \citep{2019EPJA...55...97T}. The latter may represent the widest possible domain for respective neutron star observables to be consistent with the low density input from modern calculations of neutron-rich matter based on chiral effective field theory, and also include strong phase transitions which could lead to drastic change in the stiffness of EOSs. 
\vskip +16mm
%--------|---------|---------|---------|---------|---------|---------|---------|
\section{Hybrid star structure and tidal deformability} 
\label{sec:l}

%__________________________________________________________________
\subsection{Topology of the mass-radius relation}

A very important constraint to be fulfilled is the maximum mass of neutron stars supported by different EOSs, which has to be compatible with the observational data. 
We show in Fig.~\ref{fig:rcontour} on the $(p_{\rm trans},\Delta\varepsilon/\varepsilon_{\rm trans})$ plane the contours of the maximum mass of hybrid stars $M_{\rm max}$ as well as the radius of the maximum-mass stars $R_{\rm max}$, with $L=30-60$ MeV for $c^2_{\rm QM}=1/3$ (left panel) and $c^2_{\rm QM}=1$ (right panel). 
The region inside the $M_{\rm max}=2\,M_{\odot}$ contours (red solid and red dashed) corresponds to EOSs for which the maximum mass is below $2\,M_{\odot}$, and therefore are considered excluded by the observation of stars with masses $\sim 2\,M_{\odot}$~\citep{2010Natur.467.1081D,2016ApJ...832..167F,2018ApJS..235...37A,2013Sci...340..448A,2020NatAs...4...72C}.
We also add $M_{\rm max}=2.14\,M_{\odot}$ contours (green solid and green dashed) in the right panel ($c^2_{\rm QM}=1$), corresponding to the central value of the heaviest pulsar recently discovered MSP J0740+6620~\citep{2020NatAs...4...72C}. The excluded region would be larger if more massive stars were to be observed.

For high-density EOSs with $c^2_{\rm QM}=1$, the excluded region is most limited, which allows a reasonable range of transition pressures and energy density discontinuities that are compatible with the observation. However, for high-density matter with $c^2_{\rm QM}=1/3$, the $M_{\rm max}\geq2\,M_{\odot}$ constraint eliminates almost the entire CSS parameter space. Besides, decreasing the stiffness of the nuclear matter EOS from the $L=60$ MeV (solid curves) to $L=30$ MeV (dashed curves) also enlarges the excluded region. Nevertheless, differences among nuclear matter EOSs generally induce less significant effects compared to those in quark matter EOS.  

There exists a critical value for the energy density jump $\Delta\varepsilon$ (black solid lines in Fig.~\ref{fig:rcontour}) below which a stable hybrid star branch connected to the hadronic star branch should be present ~\citep{1971SvA....15..347S,1983A&A...126..121S,1998PhRvD..58b4008L}, 
\begin{eqnarray}
\frac{\Delta\varepsilon_{\rm crit}}{\varepsilon_{\rm trans}} = \frac{1}{2} + \frac{3}{2}\frac{p_{\rm trans}}{\varepsilon_{\rm trans}}\ ,
\label{eq:limit}
\end{eqnarray}
which was obtained by performing an expansion in powers of the size of the core of high-density phase, in the presence of a sharp discontinuity in the energy density.
For energy density discontinuities above the critical value, the sequence of stars will become unstable immediately after the central pressure reaches above $p_{\rm trans}$.
Also in Fig.~\ref{fig:rcontour}, regions enclosed by the black dashed curves where the disconnected hybrid star branch exists are insensitive to the details of the nuclear matter EOS, but depends significantly on the value of $c^2_{\rm QM}$; see e.g.~\citet{2013PhRvD..88h3013A}.

Following the radius contours for the maximum-mass star $R_{\rm max}$ (blue solid and blue dashed) one can search for the minimum radius for a given EOS, as the smallest hybrid star is typically the heaviest one. 
The border of the $M_{\rm max}\geq2\,M_{\odot}$ allowed region excludes those contours with $R_{\rm max}$ greater than $11.5$ km for $c^2_{\rm QM}=1/3$, and greater than $9$ km for $c^2_{\rm QM}=1$, respectively.
The most compact stars with radii as small as $9$ km occur when the high-density phase has the largest possible speed of sound $c^2_{\rm QM}=1$, with a low transition pressure ($n_{\rm trans}\le2\,n_0$) and a fairly large energy density discontinuity $\Delta\varepsilon\gtrsim\varepsilon_{\rm trans}$.

%__________________________________________________________________
\subsection{Symmetry energy effects}

%fig
\begin{figure}[htb]
\centering
\includegraphics[width=21pc]{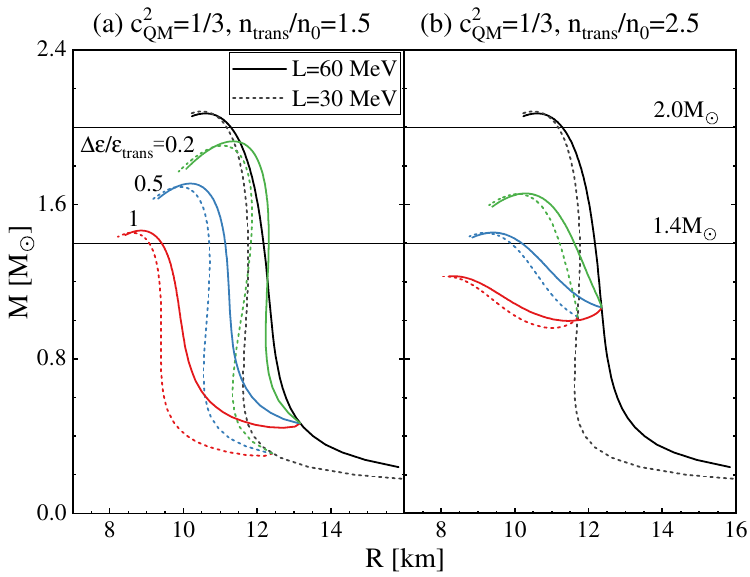}
\vskip 1mm\
\includegraphics[width=21pc]{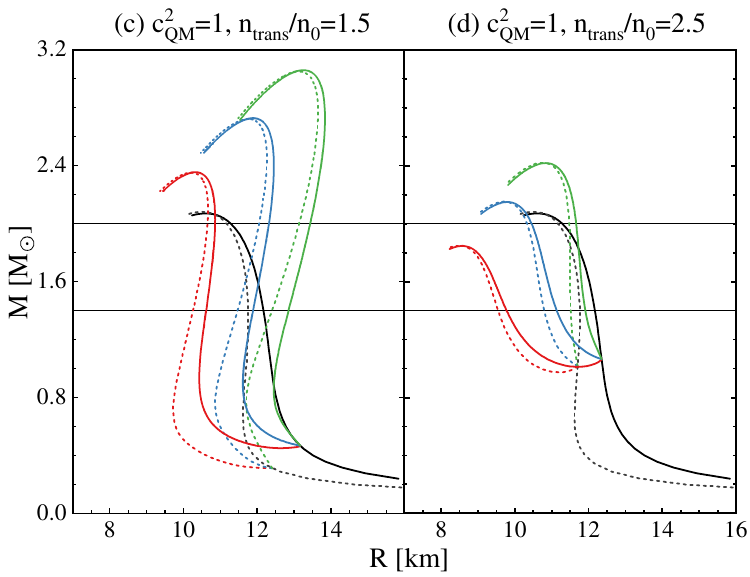}
\caption{Mass-radius relations for hybrid stars (colored curves) with transition density $n_{\rm trans}/n_0=1.5,2.5$ and different transition strengths $\Delta\varepsilon/\varepsilon_{\rm trans}=0.2,0.5,1$. Hereafter results with the symmetry energy slope $L=60~\rm MeV$ are represented by solid curves and those with $L=30$ MeV by dashed curves. The squared sound speed in quark matter is fixed to be $c^2_{\rm QM}=1/3$ (left panels) or $c^2_{\rm QM}=1$ (right panels). Purely hadronic stars are also shown for comparison (black curves). The horizontal lines in each panel indicate $M=1.4, 2.0 \, M_{\odot}$.
}
\label{fig:cs13}
\end{figure}

%fig
\begin{figure}[htb]
\centering
\vskip 2mm
\includegraphics[width=20.5pc]{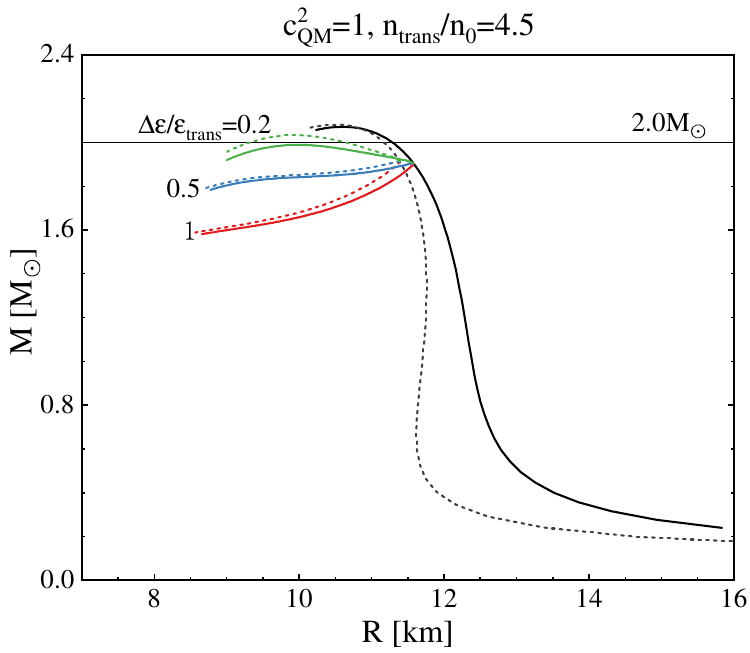}
\vskip-4mm\
\caption{Mass-radius relations for hybrid stars (colored curves) with fixed transition density $n_{\rm trans}/n_0=4.5$ and symmetry energy slope values $L=30~\rm MeV$ (dashed), $60~\rm MeV$ (solid). The squared sound speed in quark matter is fixed to be $c^2_{\rm QM} =1$. Purely hadronic stars are also shown for comparison (black curves). 
}
\label{fig:L45}
\end{figure}

In Fig~\ref{fig:cs13}, we explicitly show the mass-radius relation for hybrid stars with various transition densities $n_{\rm trans}/n_0=1.5,2.5$ and energy density discontinuities $\Delta\varepsilon/\varepsilon_{\rm trans}=0.2,0.5,1$, for $c^2_{\rm QM}=1$ and $1/3$, respectively. 
We confirm that increasing the transition density and/or the energy density discontinuity decreases the stellar mass, and thus heavy hybrid stars can be achieved by applying low transition density with small energy density discontinuity. Since the QMF EOS for hadronic matter is relatively soft, a large region of the ($n_{\rm trans}/n_0, \Delta\varepsilon/\varepsilon_{\rm trans}$) parameter space should be considered eliminated by $\sim 2\,M_{\odot}$ pulsar observations, especially for the soft $c^2_{\rm QM}=1/3$ case; see Fig.~\ref{fig:rcontour}.

In the same figure, we also illustrate how the mass-radius relation would be modified by varying the symmetry energy slope parameter $L$ from 30 to 60 MeV. As previously studied in the QMF model without hadron-quark phase transition~\citep{2018ApJ...862...98Z}, the radius of the maximum-mass star $R_{\rm max}$ is only slightly affected by the $L$ value~\citep[e.g.,][]{2001ApJ...550..426L,2006PhLB..642..436L}; we find that the conclusion still holds true in the presence of phase transition. This is mainly because the primary factor that determines maximum-mass star properties is the stiffness in the high-density quark phase. Nevertheless, change in the radius of the maximum-mass star due to variation in $L$ is relatively more evident for lower transition density $n_{\rm trans} $ and smaller energy density discontinuity $\Delta\varepsilon$.

From \citet{2018ApJ...862...98Z}, the radii of a $1.4 \,M_{\odot}$ hadronic star in QMF models are $R_{1.4}=11.76$ km and $R_{1.4}=12.17$ km, for $L=30$ MeV and $L=60$ MeV, respectively, with central density $\approx 3.1\,n_0$ and a relative difference of $\approx3.5\%$. For hybrid stars with $n_{\rm trans}/n_0=1.5$ and $\Delta\varepsilon/\varepsilon_{\rm trans}=0.2$, the corresponding values are $R_{1.4}=11.83$ km and $R_{1.4}=12.31$ km for $c^2_{\rm QM}=1/3$ with central density $\approx 2.3\,n_0$, and $R_{1.4}=12.43$ km and $R_{1.4}=12.85$ km for $c^2_{\rm QM}=1$ with central density $\approx 1.7\,n_0$, and a similar relative difference. 
Normally a strong correlation between $L$ and $R_{1.4}$ has been observed for standard extrapolation of the EOS from saturation (where $L$ is defined) to higher densities (e.g. $2-3\,n_{0}$ reached by the center of $1.4\,M_{\odot}$ stars). Nevertheless, given many parameters to tune and adjust in different models, sometimes the correlation breaks down especially when effective masses are varied in different context. Such possibilities have been discussed in \citet{2018PhRvC..98f5804H} \& \citet{2019PhRvD.100j3022H}. 

In contrast, it is worth mentioning that the role of $L$ is studied in a unified and consistent manner in the present work with all other saturation properties fixed at their empirical values, such as the saturation density, the binding energy, the incompressibility, nucleon effective mass, as well as the symmetry energy at saturation. Consequently, the differences among the nuclear matter EOSs are dominated by the differences in their symmetry energy slope values.
Future radius measurements of canonical-mass stars $\sim 1.4\, M_{\odot}$ with better accuracy that might distinguish these relative differences would help improve the uncertainty analysis of nuclear matter parameters such as variations in $L$. Along this line, a recent study extended symmetry expansion to higher densities, and pointed out that the radii of heavy stars might carry important information on the high-density behavior of nuclear symmetry energy~\citep{2020ApJ...899....4X}.

The representative mass-radius relations shown in Fig.~\ref{fig:cs13} demonstrate the topology of a hybrid branch connected or disconnected to the normal hadronic branch, as depicted in details in \citet{2013PhRvD..88h3013A}. With sufficiently high transition density and large energy density discontinuity (i.e. upper-right corners of the contour plots in Fig.~\ref{fig:rcontour}), no stable hybrid star branch exists; see also in Fig.~\ref{fig:L45} for $\Delta\varepsilon/\varepsilon_{\rm trans}=0.5,1$. In such cases, effect from the  slope parameter $L$ is rather limited because only hadronic stars are stable. We find that if the hadron-quark phase transition takes place above $n_{\rm trans}\gtrsim 4\,n_0$, it is in general difficult to derive further reliable constraints on the nuclear matter parameters. 
In fact, a recent analysis using the NJL model for the quark matter found that the density gap should occur for small values $n_{\rm trans}\le 4\,n_0$ in order to sustain a considerable quark core size~\citep{2020PhRvD.101l3030F}.

%fig
\begin{figure}%[htb]
\centering
\includegraphics[width=20.5pc]{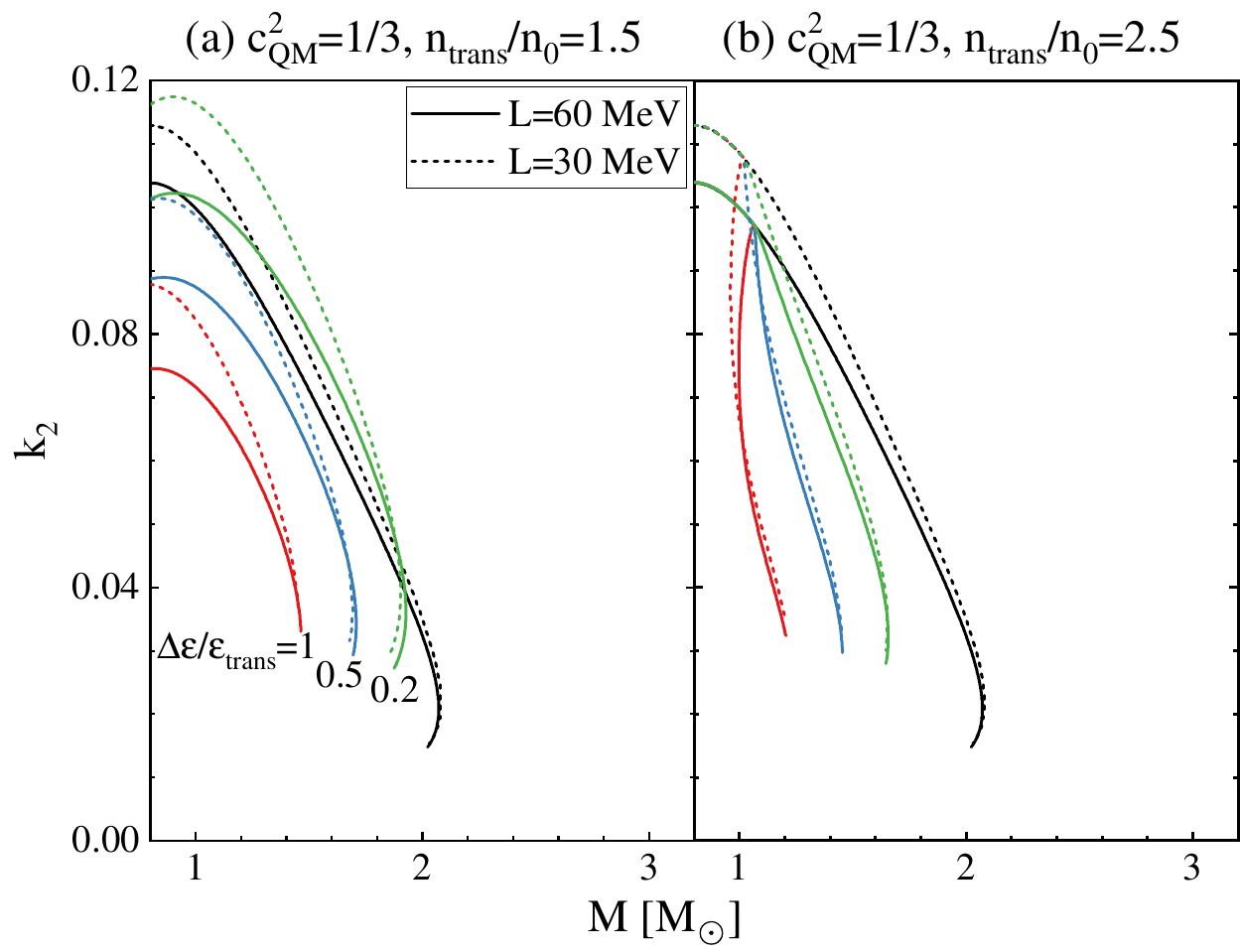}
\vskip 2mm\
\includegraphics[width=20.5pc]{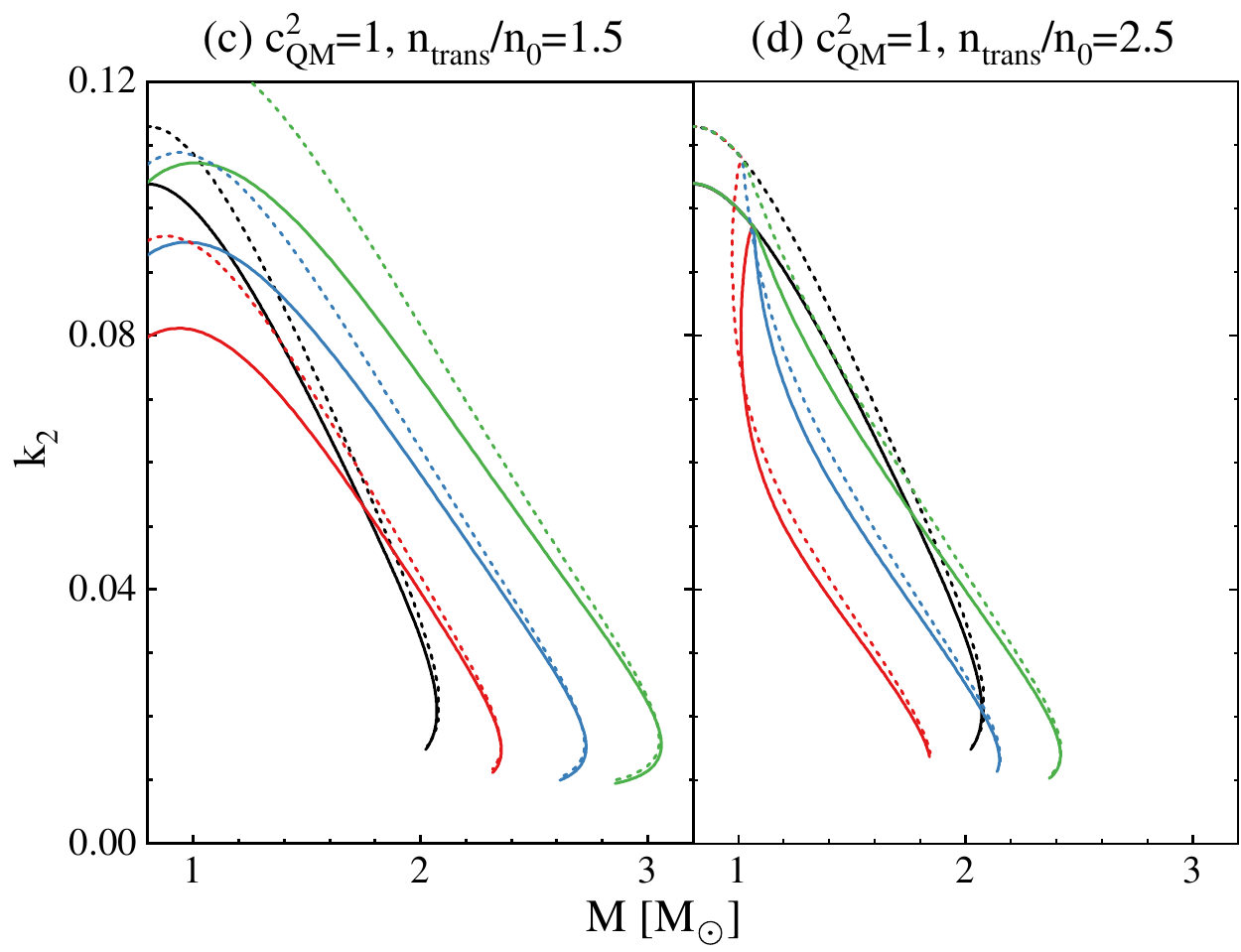}
\caption{Love number $k_2$ vs. mass for hybrid stars (colored curves) with fixed transition density $n_{\rm trans}/n_0=1.5,2.5$ and different transition strengths $\Delta\varepsilon/\varepsilon_{\rm trans}=0.2,0.5,1$; the symmetry energy slope values are $L=60~\rm MeV$ (solid curves) and $L=30$ MeV (dashed curves). The squared sound speed in quark matter is fixed to be $c^2_{\rm QM}=1/3$ (upper panels) or $c^2_{\rm QM}=1$ (lower panels). Purely hadronic star results are also shown for comparison (black curves).}\label{fig:l_k2} 
\end{figure}

%__________________________________________________________________
\subsection{Tidal deformability} 

In a coalescing neutron star binary, changes in the orbital phasing due to the components’ mutual tidal interaction leave a detectable imprint in the gravitational wave signal, and the measured tidal deformabilities can then inform constraints on the neutron star EOS.
How easily the bulk matter in a star is deformed by an external tidal field is encoded in the tidal Love number $k_2$, the ratio of the induced quadruple moment $Q_{ij}$ to the applied tidal field $E_{ij}$~\citep{2009PhRvD..80h4035D,1992PhRvD..45.1017D,2008ApJ...677.1216H},
$Q_{ij}=-k_2\frac{2R^5}{3G}E_{ij}$,
where $R$ is the neutron star radius. 
The dimensionless tidal deformability $\Lambda$ is related to the compactness $M/R$ and the Love number $k_2$ through $\Lambda = \frac{2}{3}k_2(M/R)^{-5}$. 
It is noteworthy that the computation on the tidal deformability requires additional treatment for sharp phase transitions with a finite discontinuity in the  energy density~\citep{2009PhRvD..80h4035D,2010PhRvD..82b4016P,Hinderer10}. 
We have checked that tidal deformability results for hybrid stars within the QMF model obey the I-Love-Q universal relation~\citep{2013Sci...341..365Y,2017PhR...681....1Y} reasonably well, with errors less than $2\%$ which is in agreement with previous works (see e.g.,  \citet{2019PhRvD..99h3016C}). 
The mass-weighed tidal deformability $\tilde{\Lambda}$ of a binary system
\begin{eqnarray}
\tilde{\Lambda} = \frac{16}{13}\frac{(m_1+12m_2)m_1^4}{(m_1+m_2)^5}\Lambda_1 + (1\leftrightarrow2),
\end{eqnarray}
as a function of the chirp mass $\mathcal{M}=(m_1m_2)^{3/5}/(m_1+m_2)^{1/5}$, can be accurately measured during the inspiral, and is relatively insensitive to the
unknown mass ratio $q = m_2/m_1$ ($m_1$ and $m_2$ are the masses of the components)~\citep[e.g.,][]{2018ApJ...852L..29R,2018ApJ...857L..23R,2019PhRvD..99d3010C}.

%fig
\begin{figure}%[htb]
\centering
\includegraphics[width=20.5pc]{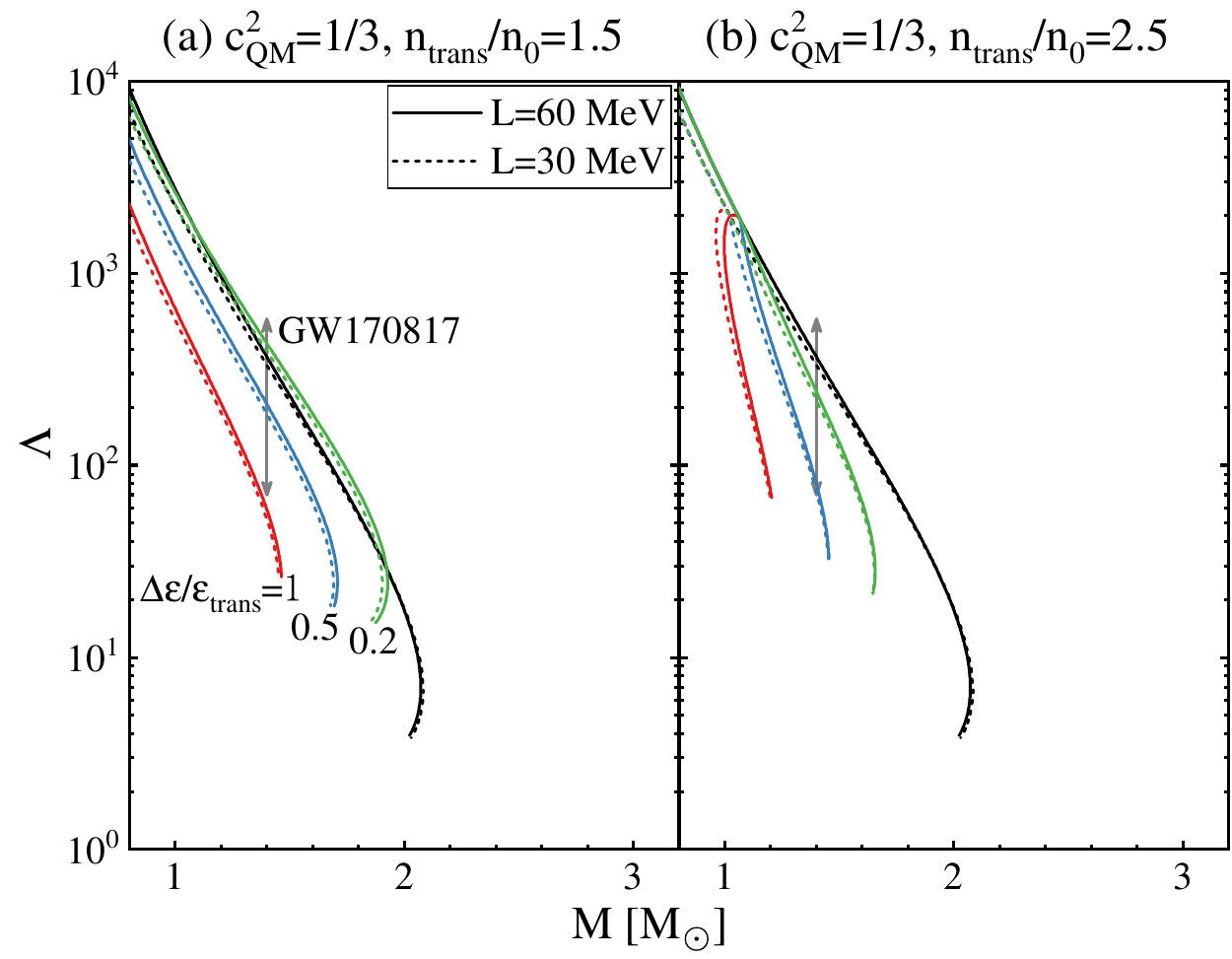}
\vskip 0.8mm\
\includegraphics[width=20.5pc]{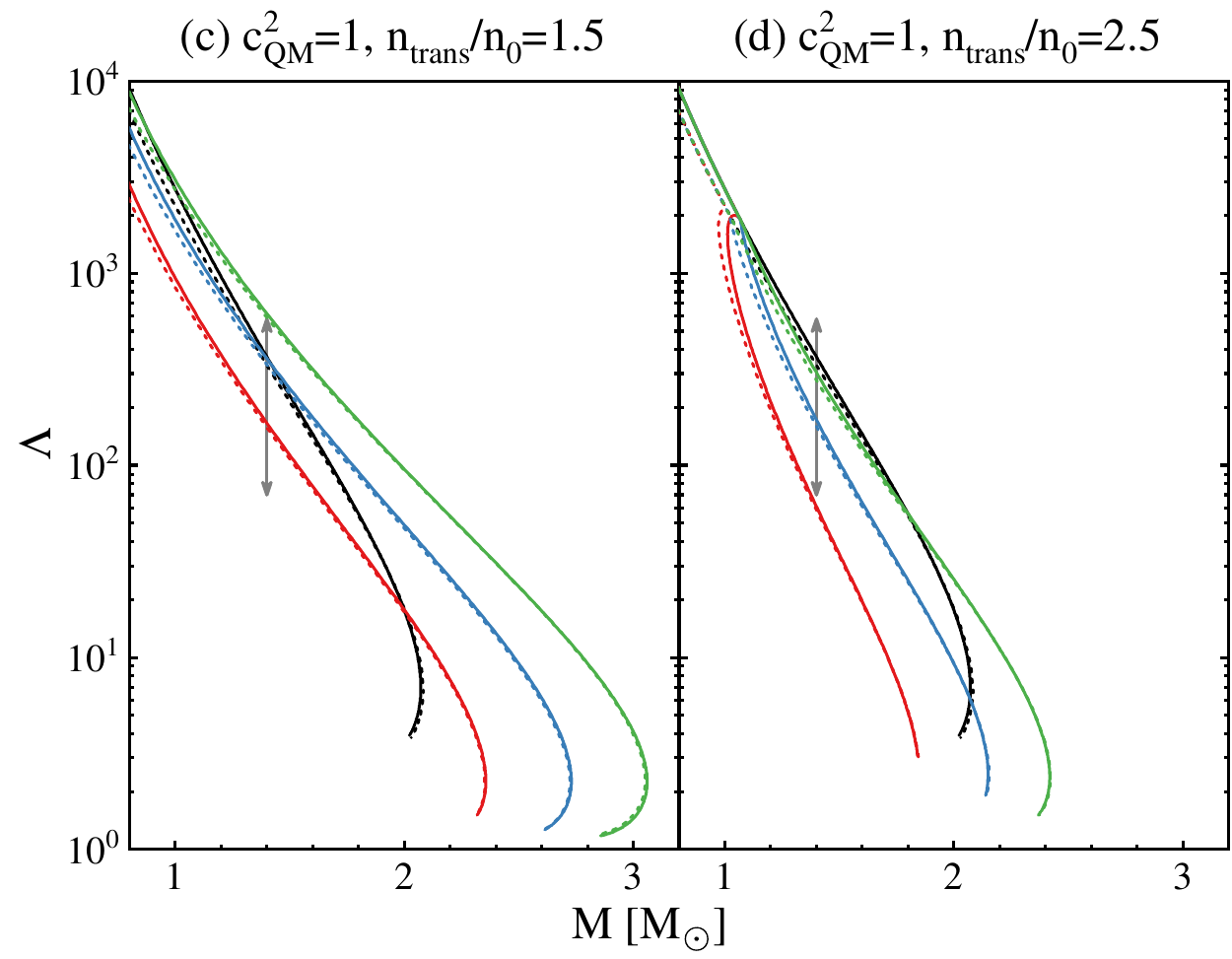}
\caption{Same as Fig.~\ref{fig:l_k2}, but showing the dimensionless tidal deformability $\Lambda$. Also displayed is the LIGO/Virgo constraint~\citep{2018PhRvL.121p1101A} from GW170817 on the tidal deformability for $1.4 \,M_{\odot}$ stars $\left(\Lambda_{\rm 1.4}=190^{+390}_{-120}\right)$, using the PhenomPNRT waveform model at a 90\% confidence level. Note that this constraint was derived without taking into account possible phase transitions.}
\label{fig:l_lam} 
\end{figure}

%fig
\begin{figure*}[htb]
\centering
\includegraphics[width=20.0pc]{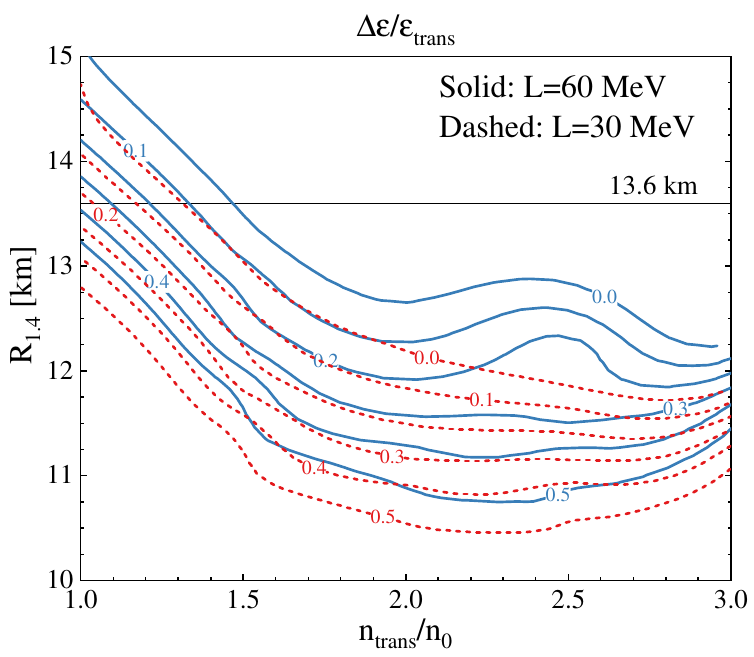}
\includegraphics[width=20.0pc]{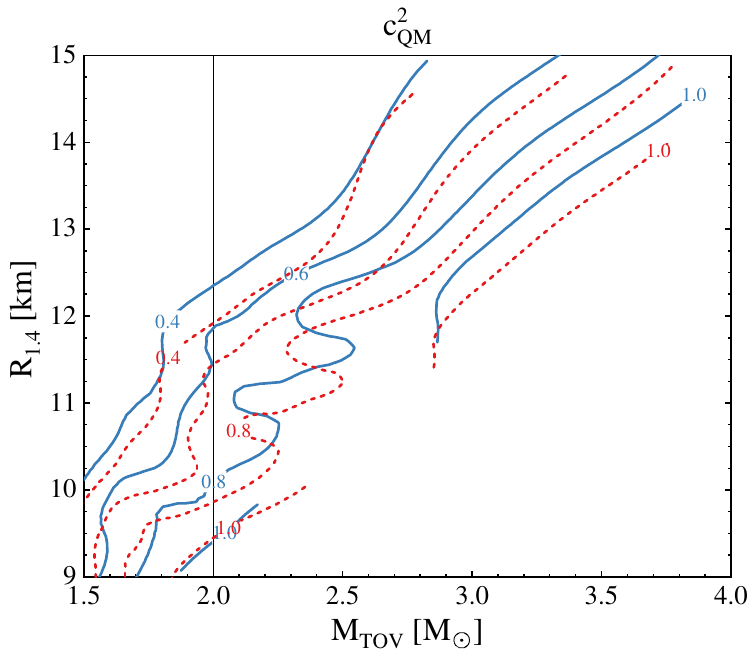}
\caption{Left: Radius of a $1.4\,M_{\odot}$ hybrid star vs. the transition density, with the red dashed (blue solid) curves showing the energy density discontinuity $\Delta\varepsilon/\varepsilon_{\rm trans}$ contours for $L=30$ MeV ($60$ MeV); the horizontal line indicates an upper bound $R=13.6$~km consistent with recent observations. Right: Radius of a $1.4 \,M_{\odot}$ hybrid star vs. the maximum mass, with the red dashed (blue solid) curves showing the squared sound speed $c^2_{\rm QM}$ contours for $L=30$ MeV ($60$ MeV). 
There are cases for which no $1.4 \,M_{\odot}$ hybrid star is possible shown with breaks in the curves. The vertical line marks the lower bound on the maximum mass $2\,M_{\odot}$. 
}
\label{fig:R14}
\end{figure*}

%fig
\begin{figure}%[htb]
\centering
\includegraphics[width=20.0pc]{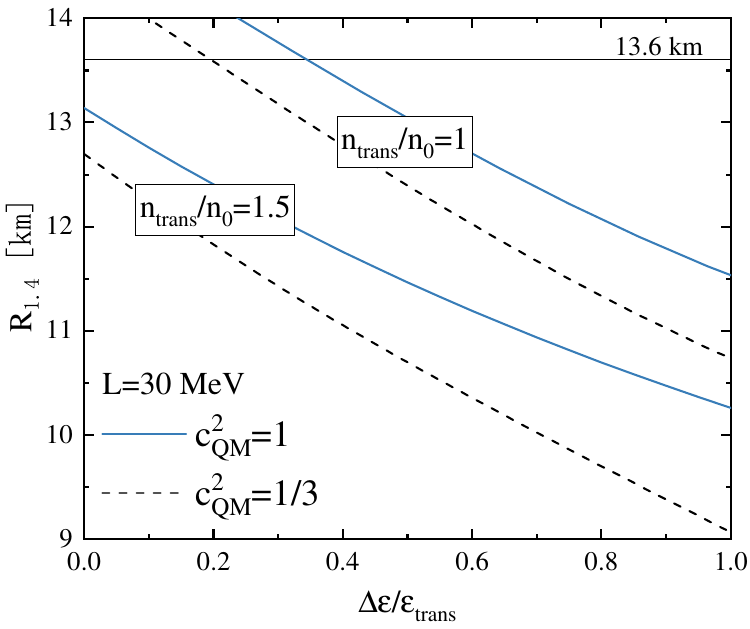}
\caption{Radius of a $1.4\,M_{\odot}$ hybrid star vs. the energy density discontinuity $\Delta\varepsilon/\varepsilon_{\rm trans}$, for the transition density $n_{\rm trans}=1,\ 1.5\,n_0$. \\
}
\label{fig:R14_De}
\end{figure}

We show the effects of the symmetry energy slope $L$ on the tidal Love number $k_2$ and the tidal deformability $\Lambda$ in Fig.~\ref{fig:l_k2} and Fig.~\ref{fig:l_lam}, respectively. We mainly discuss the effect of $L$ since $k_2$ and $\Lambda$ have been found to be essentially independent of the symmetry energy $E_{\rm sym}$ itself~\citep[e.g.,][]{2018PhRvC..98c5804M,2019PhRvC.100c5801P,2019ApJ...885..121R}. 
Figure~\ref{fig:l_k2} shows that a larger $L$ leads to a smaller $k_2$ for hybrid stars, which is similar to that found in normal hadronic stars~\citep{2018ApJ...862...98Z,2019JPhG...46c4002D}, and the $L$ effects are more evident for light stars than massive stars close to the maximum mass. The conclusions remain valid for a large variation of the phase transition parameters, i.e.  transition density $n_{\rm trans}$, the energy density discontinuity $\Delta\varepsilon$, and the quark matter speed of sound $c_{\rm QM}$. 
Also, the symmetry energy slope tends to have smaller influence when the phase transition appears at higher density. 
The dependence of $\Lambda$ on $L$ is less sensitive than that of $k_2$ as can be seen in Fig.~\ref{fig:l_lam}, mainly due to the competitive role played by the factor of $R^5$: the increase of $R$ with $L$ finally weakens the decrease of $k_2$ with $L$. As a result, the tidal deformability overall is not subject to the symmetry energy effects with its slope in the range of $30-60$ MeV. Similar conclusions have been drawn in \citet{2019JPhG...46g4001K}
with EOSs constrained by heavy-ion collision data, that measuring $\Lambda$ alone may not completely determine the density dependence of the symmetry energy.

%--------|---------|---------|---------|---------|---------|---------|---------|
\section{Confronting multi-messenger observations on $M_{\rm max}$, $R_{1.4}$, $\Lambda$, and $\tilde\Lambda$
} 
\label{sec:14}

\begin{figure}%[htb]
\centering
\includegraphics[width=20.0pc]{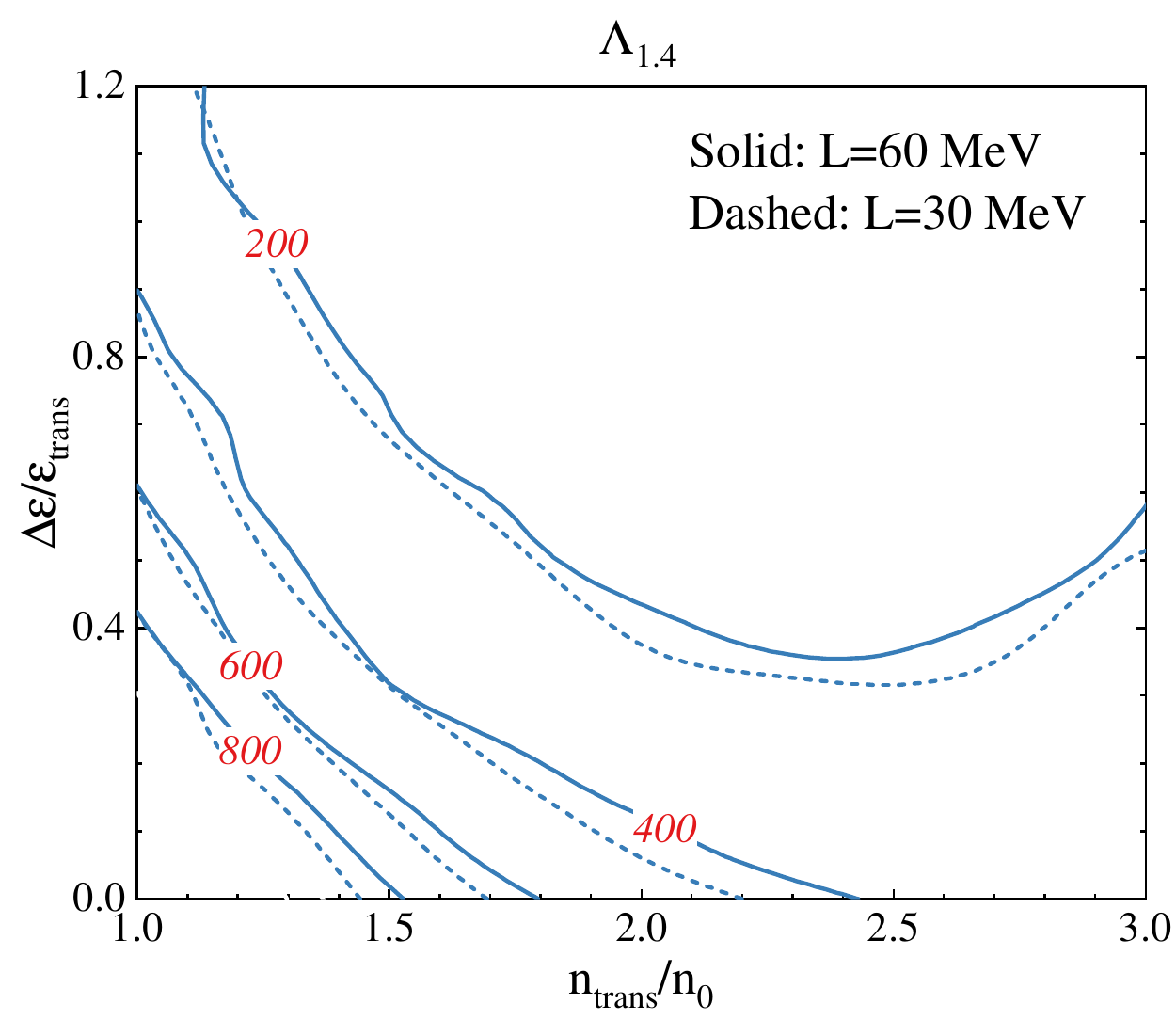}
\caption{Contour plots showing the tidal deformability of a $1.4 \,M_{\odot}$ hybrid star $\Lambda_{1.4}$ as a function of the CSS parameters of the high-density EOS. The dependence on the CSS parameters ($n_{\rm trans}/n_0,\Delta\varepsilon/\varepsilon_{\rm trans}$) are shown with the quark matter sound speed varying between $c^2_{\rm QM} =1/3$ and $1$ and the symmetry energy slope being $L=30~\rm MeV$ (dashed curves) and $L=60~\rm MeV$ (solid curves). 
}
\label{fig:Lam14}
\end{figure}

\begin{figure}%[htb]
\centering
\includegraphics[width=20.0pc]{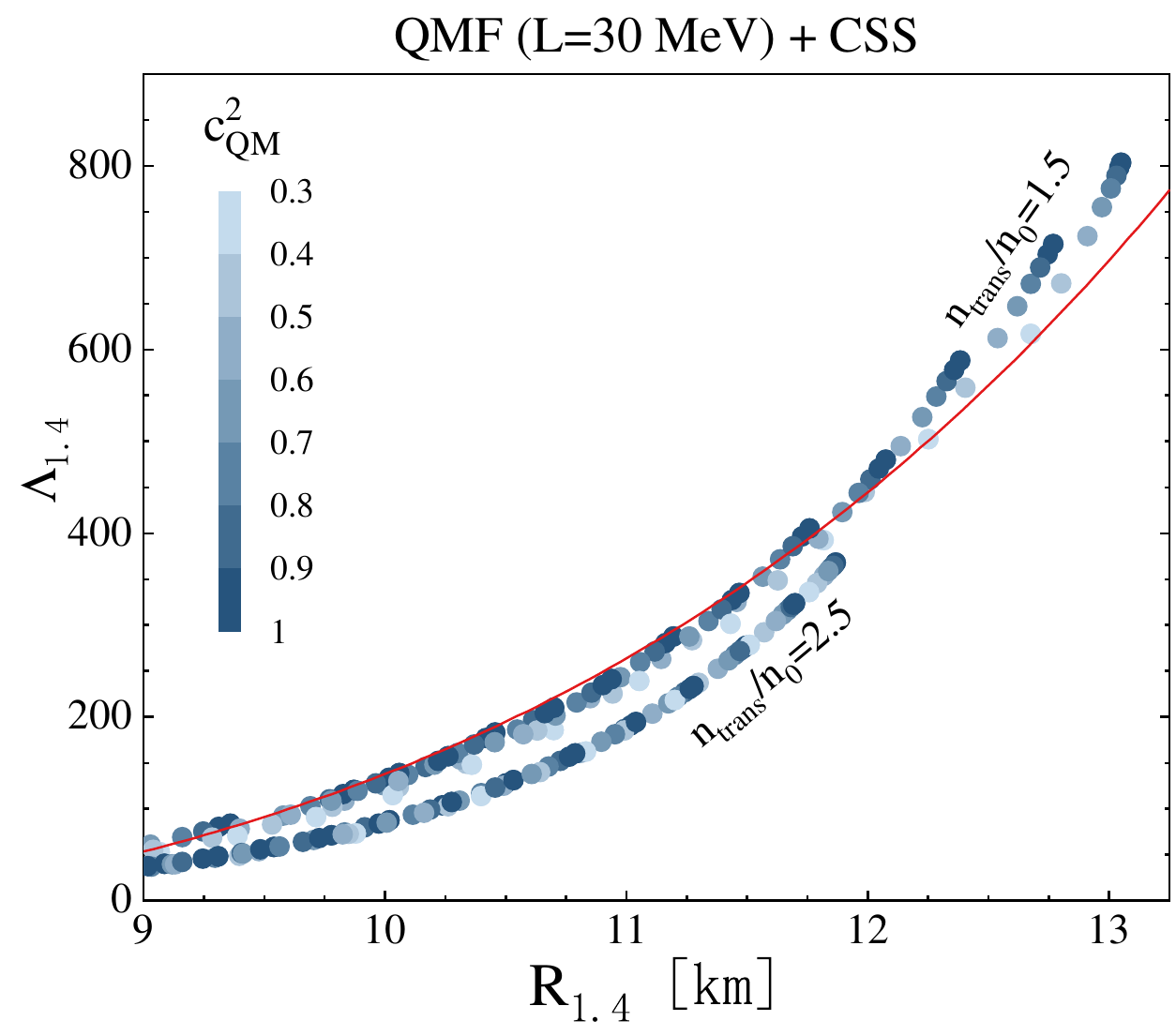}
\caption{Tidal deformability vs. radius for a $1.4 \,M_{\odot}$ hybrid star with the  transition density $n_{\rm trans}/n_0=1.5,2.5$ and the symmetry energy slope $L=30~\rm MeV$. 
The $n_{\rm trans}/n_0=2.5$ curve is shorter than the $n_{\rm trans}/n_0=1.5$ case due to greater softening of the hybrid star EOSs and consequently lower tidal deformability.
The squared sound speed is explicitly indicated varying between $c^2_{\rm QM} =1/3$ and $1$.
The energy density discontinuity is calculated up to $\Delta\varepsilon/\varepsilon_{\rm trans}=1.5$. 
For different transition densities, there appear universal relations for a given neutron star mass between $\Lambda$ and $R$~\citep[e.g.,][]{2018PhRvL.120q2703A,2019EPJA...55...97T} in the case of no phase transitions~\citep[e.g.,][]{2017PhR...681....1Y,2018PhRvL.120q2702F,2018PhRvL.121i1102D,2018ApJ...857L..23R,2018PhRvC..98c5804M,2019PhRvC.100c5801P,2019PhRvD..99l1301Z}. The universal relation for $1.4 \,M_{\odot}$ compact stars from~\citet{2019EPJA...55...97T} is also shown for comparison.}
\label{fig:LamR}
\end{figure}

\begin{figure*}%[htb]
\centering
\includegraphics[width=21.2pc]{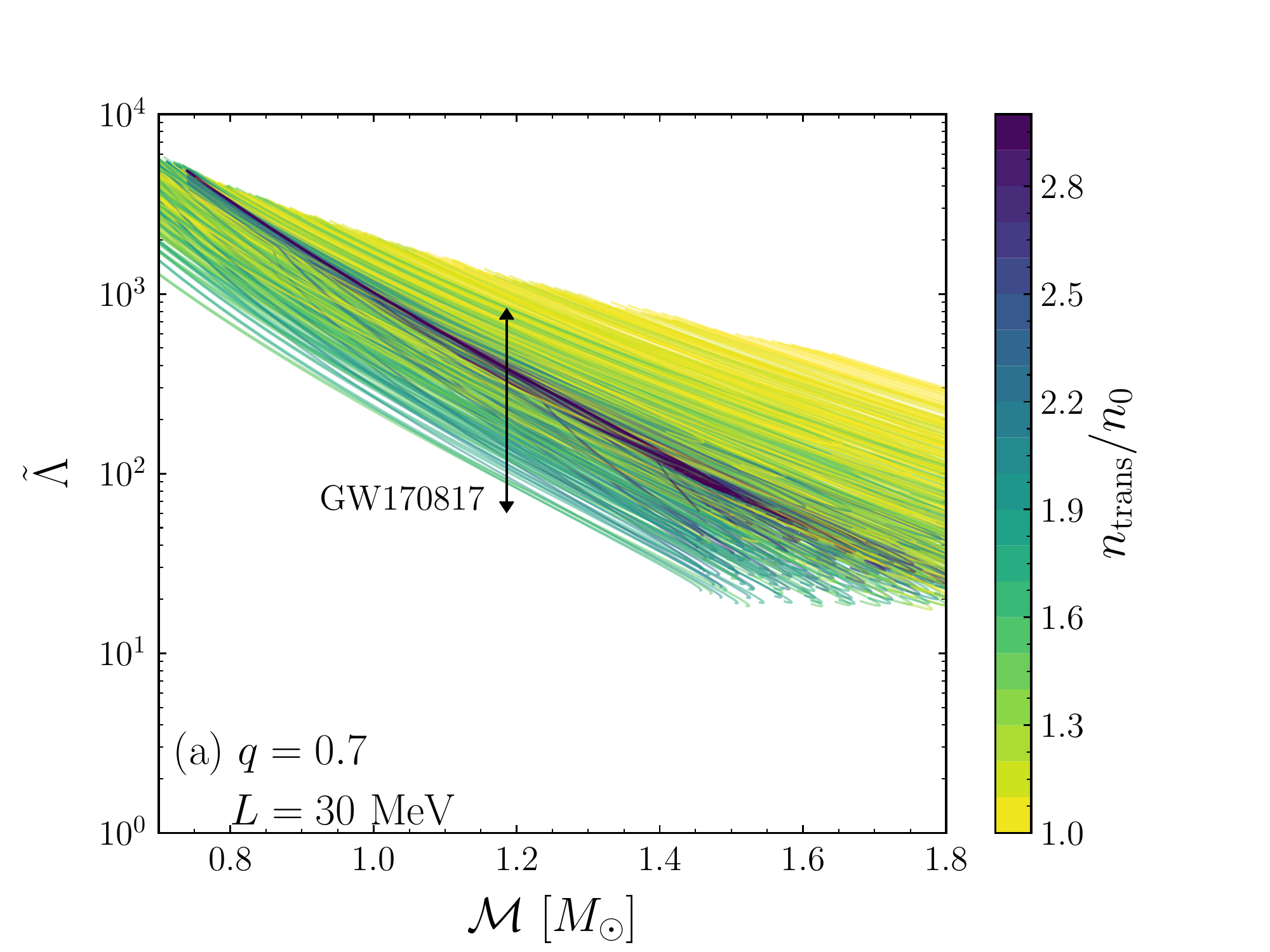}
\includegraphics[width=21.2pc]{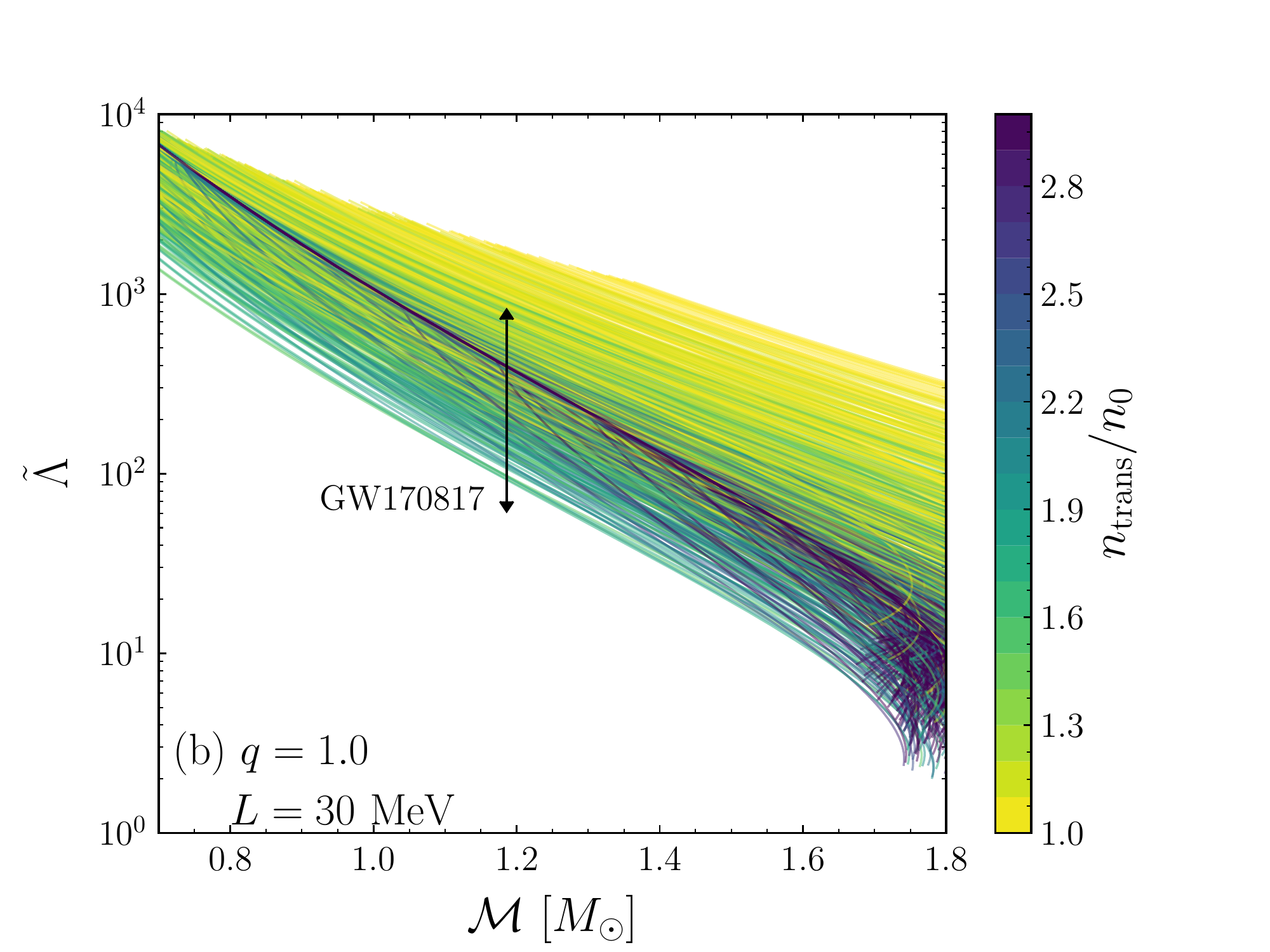}
\caption{Current uncertainties in the combined tidal deformability $\tilde{\Lambda}$ for hybrid stars as a function of the chirp mass $\mathcal{M}$, depending on phase transition parameters ($n_{\rm trans}/n_0,\Delta\varepsilon/\varepsilon_{\rm trans}, c^2_{\rm QM}$).
The transition density $n_{\rm trans}/n_0$ is explicitly indicated.
The calculations are done for the symmetry energy slope $L=30~\rm MeV$. 
The mass ratio is chosen to be $q=0.7$ (left) and $q=1$ (right). The chirp mass for GW170817 $\mathcal{M}=1.186 \,M_{\odot}$ is also indicated with the constraint of $70\leq\tilde{\Lambda}\leq 720$~\citep{2019PhRvX...9a1001A}.
We note that very small values of $\tilde{\Lambda}\lesssim 20 $ for high chirp masses $\mathcal{M}\sim 1.6-1.8 \,M_{\odot}$ are only possible for $q=1$, not allowed for $q = 0.7$; see text for details.
}
\label{fig:Lam}
\end{figure*}

Systematically, we carry on with calculations for the mass-radius of hybrid stars spanning the whole parameter space of the speed of sound, with the transition density up to $n_{\rm trans}=6\,n_0$ and the energy density discontinuity up to $\Delta\varepsilon=1.5\,\varepsilon_{\rm trans}$. The calculations are performed using two values of the symmetry energy slope parameter $L=30$ MeV and $L=60$ MeV, and the results are shown in Figs.~\ref{fig:R14}-\ref{fig:Lam}.

Figure~\ref{fig:R14} displays the correlation of the radius of a $1.4 \,M_{\odot}$ hybrid star $R_{1.4}$ with the transition density $n_{\rm trans}/n_0$ (left panel) and with the maximum mass $M_{\rm max}$ (right panel). In general, there exists an anti-correlation between $R_{1.4}$ and $n_{\rm trans}/n_0$, and a correlation between $R_{1.4}$ and $M_{\rm max}$.
A conservative upper limit of $13.6$ km for $R_{1.4}$ can be obtained with different analyses~\citep[e.g.,][]{2016PhRvC..94c5804F,2017ApJ...850L..34B,2018PhRvL.121p1101A,2018PhRvL.120q2703A,2018ApJ...860..139B,2018PhRvL.121i1102D,2018PhRvL.120z1103M, 2018PhRvC..98d5804T,2019PhRvD..99j3009M,2020ApJ...893L..21R}, 
and since our hadronic EoSs tend to have a $1.4\,M_{\odot}$ hadronic star well suited for this bound ($R_{1.4}^{\rm had}\approx 12$~km), we primarily apply this condition to hybrid stars with $1.4 \,M_{\odot}$ and constrain phase transition parameters accordingly.\footnote{Note that certain disconnected hybrid configurations can have the hadronic branch violating $R_{1.4}^{\rm had}<13.6$~km, which our QMF model does not present.}

In the left panel, the upper limit of $13.6$ km for $R_{1.4}$ indicates that 
a very weak phase transition (i.e. small energy discontinuity $\Delta\varepsilon$) taking place at too low densities ($n_{\rm trans}\lesssim 1.31\,n_0$ for $L=30$ MeV or $n_{\rm trans}\lesssim 1.46\,n_0$ for $L=60$ MeV) is strongly disfavored. To illustrate, we depict in detail in Fig.~\ref{fig:R14_De} the decreasing trend of $R_{1.4}$ with $\Delta\varepsilon/\varepsilon_{\rm trans}$, with two cases of transition densities around $n_{\rm trans}=1.31\,n_0$ for $L=30$ MeV (the trend for $L=60$ MeV is similar). 
On the other hand, given mass measurements of heavy pulsars one can set lower limits on $R_{1.4}$ by making use of the $R_{1.4}-M_{\rm max}$ correlation in the right panel: $M_{\rm max}\geq2\,M_{\odot}$ infers a similar lower limit of $\approx 9.6$ km on $R_{1.4}$, and with the more stringent $2.14 \,M_{\odot}$ constraint, this limit is raised slightly to $\approx 9.7$ km. 
These values are in good concurrence with other analyses in the literature based on x-ray observations or LIGO/Virgo measurements~\citep[e.g.,][]{2016EPJA...52...18S,2017ApJ...850L..34B,2018PhRvL.121p1101A, 2018PhRvL.120z1103M,2018PhRvC..98d5804T,2019ApJ...872L..16K,2019PhRvD..99j3009M,2020PhRvD.101j3029O,2020ApJ...893L..21R}, as well as theoretical predictions by previous studies of stable hybrid stars~\citep{2016EPJA...52...62A}. 
Recent analysis of NICER x-ray timing data on PSR J0030+0451 suggests $R\approx 13~\rm km$ for a $\sim1.4 \,M_{\odot}$ star~\citep{2019ApJ...887L..24M,2019ApJ...887L..21R,2019ApJ...887L..22R} based on EOSs without a phase transition, while confronting binary neutron star simulations with gravitational-wave observations obtains $R_{1.4}\approx 11~\rm km$~\citep{2020NatAs...4..625C} assuming the description in terms of nuclear degrees of freedom remains valid up to $2\,n_0$. Due to the relatively small hadrnoic $R$ realized in QMF models, our results do not account for the specific  ``third-family'' scenario that PSR J0030+0451 measured by NICER is a hadronic star with radius much larger than the more compact hybrid stars when a strong phase transition occurs above its measured mass, leading to an upper limit on the onset density $\sim1.7\,n_0$ subject to the hadronic models employed~\citep{2020ApJ...894L...8C}.

%fig

An upper limit on the maximum mass can also be indicated from $R_{1.4}<13.6$ km, which is $ M_{\rm max} < 3.6 \,M_{\odot}$. We note here that the maximum mass being almost $4 \,M_{\odot}$ from \citet{2019EPJA...55...97T}, or similarly $3.9 \,M_{\odot}$ from \citet{1996ApJ...470L..61K}, would indicate a transition density as low as $n_{\rm trans}\sim n_0$. Previously, the extreme causal equation of $P=(\varepsilon-4.6\times10^{14})c^2+P_m$ matched smoothly (i.e. without a sharp discontinuity in $\varepsilon$) to a realistic nuclear matter EOS \citep{1973NuPhA.207..298N} at about twice saturation density $n_{\rm trans}=2\,n_0$ ($P_m$ is a constant determined from the matching) was shown to result in an upper limit of $\approx3.2\,M_{\odot}$ on the maximum mass~\citep{1974PhRvL..32..324R}. Later by lowering the transition density to nuclear saturation density $n_0$, the authors found $\approx4.8\,M_{\odot}$ as an upper limit on the maximum mass~\citep{1976Natur.259..377B}.
Related discussions can also be found in e.g. \citet{2019EPJA...55...39Z}. 

These high theoretical limits on the maximum mass of neutron stars around $\approx 3-4\, M_{\odot}$ are quite beyond the observational bound of pulsars around $2.2 \,M_{\odot}$~\citep{2020NatAs...4...72C}; observations of accreting black holes, on the other hand, hinted a paucity of sources with masses below $5\, M_{\odot}$~\citep[e.g.,][]{1998ApJ...499..367B,2010ApJ...725.1918O,2011ApJ...741..103F,2012ApJ...757...36K}. 
However, binary merger involving one or two companions with masses that fall into the so-called mass gap range ($\approx3-5\, M_{\odot}$) are hard to distinguish~\citep[e.g.,][]{2020A&A...636A..20W,2020PhRvL.124g1101T,2020ApJ...892L...3A}.

In Fig.~\ref{fig:Lam14} we show contour plots of the tidal deformability for $1.4 \,M_{\odot}$ hybrid stars, $\Lambda_{1.4}$, as a function of the CSS parameters ($n_{\rm trans}/n_0,\Delta\varepsilon/\varepsilon_{\rm trans}$) of the high-density EOS. The calculations are done with the quark matter sound speed varying between $c^2_{\rm QM} =1/3$ and $1$. There are relatively small differences between the results with two different symmetry energy slope values chosen, $L=30$ MeV and $60~\rm MeV$; see also Fig.~\ref{fig:l_lam}.

%fig

%fig

%fig
\begin{figure}%[htb]
\centering
\includegraphics[width=20.0pc]{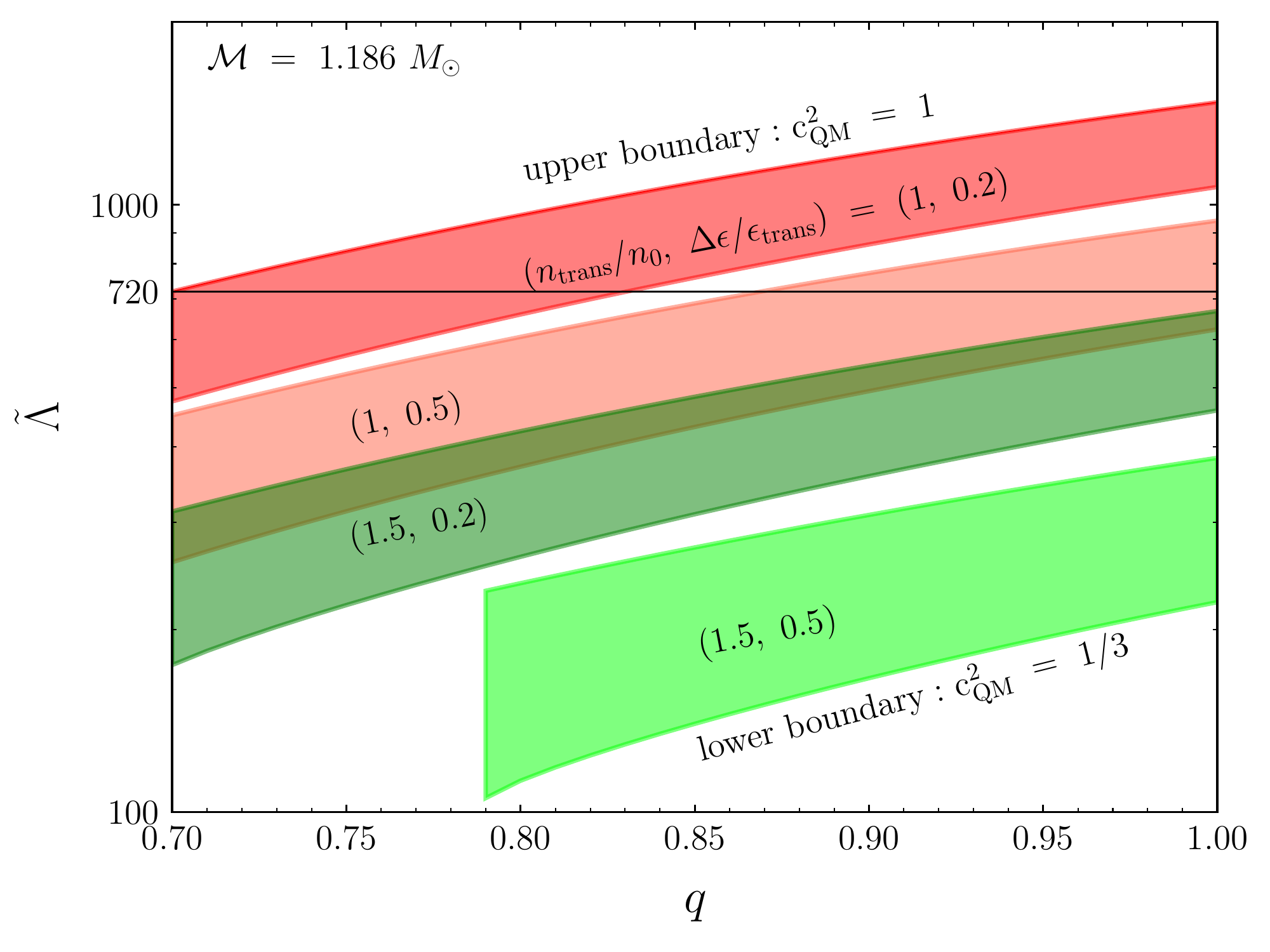}
\caption{Combined tidal deformability $\tilde{\Lambda}$ for hybrid stars as a function of the mass ratio $q$,  with the transition density $n_{\rm trans}/n_0=1,~1.5$, $\Delta\varepsilon/\varepsilon_{\rm trans}=0.2~(0.5)$ and $c^2_{\rm QM}=1/3,~1$. 
The chirp mass is fixed to be $\mathcal{M} =1.186 \,M_{\odot}$ as in GW170817, and its upper $\tilde{\Lambda}$ constraint is shown with a horizontal line.
The calculations are done for the symmetry energy slope $L=30~\rm MeV$. \\
}
\label{fig:Lam30}
\end{figure}

It has been widely discussed in the literature that there exists an empirical relation between the tidal deformability and radius for a fixed-mass star~\citep[e.g.,][]{2017PhR...681....1Y,2018PhRvL.120q2702F,2018PhRvL.121i1102D,2018ApJ...857L..23R,2018PhRvC..98c5804M,2019EPJA...55...97T,2019PhRvC.100c5801P,2019PhRvD..99l1301Z} which translates the $\Lambda$ measurement through gravitational wave observations into that of the radius. We show in Fig.~\ref{fig:LamR} our results for a $1.4 \,M_{\odot}$ hybrid star (when such configurations exist) with $L=30$ MeV and two exemplary transition densities $n_{\rm trans}/n_0=1.5,2.5$.
Increasing $c^2_{\rm QM}$ leads to larger values of $R_{1.4}$ and $\Lambda_{1.4}$, while increasing $\Delta\varepsilon$ does the opposite.
Large discontinuities in the energy density $\Delta\varepsilon$ are located in the lower-left corner of the plot: for $n_{\rm trans}/n_0=1.5$, $\Delta\varepsilon>\varepsilon_{\rm trans}$ indicates that $R_{1.4}<10.2$ km and $\Lambda_{1.4}<162$. 
It may still be possible to derive some similar empirical relation for hybrid EOSs relating $\Lambda_{1.4}$ and $R_{1.4}$, but for a given nuclear matter model the unknown threshold density $n_{\rm trans}$ has a non-trivial effect. Should further information on the phase transition density in dense matter be learned in the future, possibly from heavy-ion collision experiments, a better empirical relation can be evaluated for the use of coherent analyses of the dense matter EOS. 

Finally, Fig.~\ref{fig:Lam} illustrates current uncertainties in the combined tidal deformability $\tilde{\Lambda}$ for hybrid stars within the present QMF + CSS framework, depending on the phase transition parameters ($n_{\rm trans}/n_0,\Delta\varepsilon/\varepsilon_{\rm trans}, c^2_{\rm QM}$). We show the results for two mass ratios $q=0.7,~1$.
The $\tilde{\Lambda}$ uncertainty is comparable with the current GW170817 constraint, and tends to grow with the chirp mass. 
In addition, very small values of $\tilde{\Lambda}$ %\ang{($\le 20 $)} 
for high chirp masses $\mathcal{M}\sim 1.6-1.8\,M_{\odot}$ are only possible for the equal mass ratio $q=1$, but not allowed for $q = 0.7$. This is because more symmetric binary systems have higher chances for both components reach the more compact branch with small tidal deformabilities. 
As previously discussed, 
too weak first-order phase transition ($\Delta\varepsilon/\varepsilon_{\rm trans}\lesssim 0.2$) below $1.31 \,n_0$ or $1.46 \,n_0$ is strongly disfavored in the present study considering both heavy pulsar measurements $M_{\rm max}$ and radii constraints $R_{1.4}<13.6$~km. 
This is also consistent with our results for $\tilde\Lambda$, which indicate that too small transition densities are more likely to break the upper bound on $\tilde\Lambda$ from GW170817, as the density ranges probed by measuring tidal parameters of canonical-mass mergers from gravitational waves and radius inference of canonical-mass NSs from x-ray observations are the same.
We further illustrate this point in Fig.~\ref{fig:Lam30}, where the uncertainty bands of $\tilde{\Lambda}$ regarding the squared sound speed $c^2_{\rm QM}$ are shown as a function of the mass ratio $q$. The calculations are done for transition densities $n_{\rm trans}=1,~1.5\,n_0$ and $\Delta\varepsilon/\varepsilon_{\rm trans}=0.2,~0.5$ in the case of $L=30$ MeV; the chirp mass is fixed to be $\mathcal{M} =1.186 \,M_{\odot}$ as in GW170817. The parameter space ruled out by $\tilde{\Lambda}\leq 720$ is consistent with Fig.~\ref{fig:R14_De} where upper bound on $R_{1.4}\leq 13.6$~km is applied. 
Future measurements of more binray neutron star mergers with different chirp masses and mass ratios, with an accuracy of the extracted tidal deformability comparable to or better than GW170817, hold promise of reducing the uncertainties significantly.

%--------|---------|---------|---------|---------|---------|---------|---------|
\section{Summary} 
\label{sec:sum}

To understand the dependence of neutron star observables on both the nuclear symmetry energy (primarily its slope $L$) and the hadron-quark phase transition parameters, we extend our previous QMF model for nuclear matter and study hybrid star EOSs with quark matter in their dense cores. Assuming that the hadron-quark phase transition is of first order and characterized by a sharp interface, low-density hadronic matter described by QMF transforms into a high-density phase of quark matter modeled by the generic CSS parametrization, in terms of the critical density at which the transition occurs $n_{\rm trans}$, the strength of the transition $\Delta\varepsilon/\varepsilon_{\rm trans}$, and the ``stiffness'' of the high-density phase which we choose to vary between two extreme cases, $c_{\rm QM}^2=1/3$ (the conformal limit in perturbertive QCD matter; \textit{soft} EOS) and $1$ (the causality limit; \textit{stiff} EOS). Exploring vastly different combinations of these parameter values ($L$, $n_{\rm trans}/n_0, \Delta\varepsilon/\varepsilon_{\rm trans}, c^2_{\rm QM}$), we then extensively study and discuss masses, radii, and tidal deformabilities of hybrid stars obtained, and confront our results with constraints from multi-messenger observations although possible phase transitions were typically not taken into account in data analyses of those observations. 

While fixing the nuclear symmetry energy at its preferred value of $E_{\rm sym} =31~ \rm MeV$, a variation of its slope within the empirical range $L\approx 30-60~\rm MeV$ leads to a radius difference $\Delta R \approx 1~\rm km$ for a $1.4 \,M_{\odot}$ star, which holds true for both normal hadronic stars and hybrid stars in our calculations. 
We confirm that in the case of hybrid stars, the lower the transition threshold density $n_{\rm trans}$, the larger the maximum mass $M_{\rm max}$; the larger the discontinuity in energy density at the transition $\Delta\varepsilon$, the smaller the typical radius. PSR J0030+0451 could be either a normal neutron star or a hybrid star with a quark core, given the relatively large uncertainties in its radius inference.

Finally, parameter spaces for both the mass and radius are found to be much more extended for hybrid stars compared to those of purely hadronic ones. 
In particular, for normal neutron stars within QMF the typical radius $R_{1.4}$ and the maximum mass $M_{\rm max}$ remain close to $ \sim12~\rm km$ and $\sim2.1\,M_{\odot}$, respectively, whereas for hybrid stars, the radius can be in the range of $R_{1.4} \approx 9.6-13.6~\rm km$ while the maximum mass varies between $M_{\rm max}\approx 2-3.6\,M_{\odot}$. The combination of stiffness in high-density quark matter (that helps reach high masses) and the strength of phase transition (that ensures compatibility with small radius/tidal deformability of intermediate-mass stars), if suitably chosen, enhances compatibility with data.
We also find that to be consistent with available observational constraints, primarily from heavy pulsar mass measurements and typical radius estimates, phase transitions that are too-weak happening at low densities close to nuclear saturation are strongly disfavored. 

We conclude that it is possible to constrain the nuclear symmetry energy slope and the hadron-quark first-order phase transition properties coherently from mass-radius and tidal deformability measurements of neutron stars, in line with major goals of x-ray missions (e.g. NICER, eXTP) and LIGO/Virgo gravitational-wave detectors. Detailed information on the symmetry energy slope $L$ can be extracted from (especially the radius) measurements of canonical-mass $\sim 1.4 \,M_{\odot}$ stars, while more massive stars around $2 \,M_{\odot}$ probe the density range in the vicinity of possible quark deconfinement.
Future opportunities of studying dense matter EOS from gravitational wave signals of binary neutron star mergers are also quantitatively analysed. 
Loud gravitational-wave detection events and promising multi-messenger observations from these systems in the next decade would provide data with even better precision to help improve our understanding of the phase state of cold dense matter, such as prospects of constraining the onset density and transition strength for possible strong phase transition encountered in the neutron star interiors. \\
%\vskip -4mm

\acknowledgments
We are thankful to the anonymous referee for his or her beneficial comments and Andreas Bauswein, Bao-An Li, Luciano Rezzolla, and Enping Zhou for helpful discussions. 
The work was supported by the National Natural Science Foundation of China (Grant No.~11873040). S.H. is supported by the National Science Foundation, Grant PHY-1630782, and the Heising-Simons Foundation, Grant 2017-228. 

%\newpage
%--------|---------|---------|---------|---------|---------|---------|---------|

\end{document}